\documentclass[11pt,fleqn]{article}
\usepackage{amssymb}
\usepackage{hyperref}
\usepackage{times}
\usepackage{graphicx}
\usepackage{wrapfig}
\usepackage{fancyhdr}
\usepackage{amsmath}

\input{tcilatex}
\begin{document}

\title{Boundary Element Solution of Electromagnetic Fields for Non-Perfect Conductors at Low Frequencies and Thin Skin Depths}
\author{Nail~A.~Gumerov\thanks{Institute for Advaced Computer Studies (UMIACS), University of Maryland, College Park. Website:\url{http://www.umiacs.umd.edu/users/gumerov} E-mail:\url{gumerov@umiacs.umd.edu}}, Ross~N.~Adelman\thanks{Army Research Laboratory, Adelphi, MD. Website:\url{http://rossadelman.com/} E-mail:\url{radelman@gmail.com}}, and Ramani~Duraiswami\thanks{Dept. of Computer Science and UMIACS, University of Maryland, College Park. Website:\url{http://www.umiacs.umd.edu/users/ramani} E-mail:\url{ramani@umiacs.umd.edu}}}

\maketitle

\begin{abstract}
	A novel boundary element formulation for solving problems involving eddy currents in the thin skin depth approximation is developed. It is assumed that the time-harmonic magnetic field outside the scatterers can be described using the quasistatic approximation. A two-term asymptotic expansion with respect to a small parameter characterizing the skin depth is derived for the magnetic and electric fields outside and inside the scatterer, which can be extended to higher order terms if needed. The introduction of a special surface operator (the inverse surface gradient) allows the reduction of the problem complexity. A method to compute this operator is developed. The obtained formulation operates only with scalar quantities and requires computation of surface operators that are usual for boundary element (method of moments) solutions to the Laplace equation. The formulation can be accelerated using the fast multipole method. The method is much faster than solving the vector Maxwell equations. The obtained solutions are compared with the Mie solution for scattering from a sphere and the error of the solution is studied. Computations for much more complex shapes of different topologies, including for magnetic and electric field cages used in testing are also performed and discussed.
\end{abstract}

\section{Introduction}\label{intro}
Many systems of practical interest consist of conductors and dielectric materials. The modeling of time-varying electromagnetic fields in such systems is an important engineering problem where scattering from antennas, buildings, and various other objects of arbitrary shape must be computed. In many cases, the conductors can be modeled as perfect electric conductors, and this approximation is widely used. However, there are a number of situations where this approximation is invalid. When eddy currents appear due to finite conductivity (``non-perfectness'' of the conductor), which is caused by diffusion of the magnetic field inside the conductor, this must be accounted for in the modeling.

Scattering from a non-perfect conductor can be modeled using Ohm's law and Maxwell's equations for the electromagnetic fields inside and outside of the conductor, with the fields coupled together by transmission boundary conditions on the surface of the conductor \cite{Jackson1998book}. For time-harmonic electromagnetic fields, Ohm's law leads to the concept of complex electric permittivity, and the well-known boundary integral equations for Maxwell's equations \cite{Chew1999book} can be used. The complex electric permittivity of a conductor results in a complex wavenumber, and the reciprocal of the magnitude of this number defines the skin depth, $\delta $, or depth of penetration of the magnetic field inside the conductor. Several numerical challenges with the boundary integral solvers appear in this case where the skin depth is much smaller than the {\em characteristic length of the scatterer}, $a$ (i.e., $\delta \ll a$). Indeed, the size of the boundary elements, $\Delta $, should be much smaller than the skin depth (i.e., $\Delta \ll \delta $) to enable valid discrete representation of the continuous electromagnetic fields. Such a requirement may lead to huge numbers of boundary elements and drastically increase the computational complexity of the problems and even make them practically unsolvable.

Another observation is that for many practical problems (e.g., involving good conductors in low-conductivity media, such as air), the conductivity of the carrier medium is many orders of magnitude smaller than the conductivity of the conductor (e.g. conductivities of air can be 20 orders of magnitude or smaller than the conductivity of good conductors, such as aluminum). Thus, in certain frequency ranges, situations can occur where the wavelength in the carrier medium is much larger than $a$ (i.e., low frequency), while the skin depth inside the scatterer is much smaller than $a$. Moreover, this situation can arise in practically important problems (e.g. the wavelength of 1 MHz electromagnetic field in air is 300 m, while the skin depth of 1 MHz electromagnetic field in aluminum is 8.2$\times10^{-5}$ m, so for objects of size, $a \sim 10^{-3}$--$10^1$ m, we have such a case). 

The goal of this paper is to simplify Maxwell's equations  for low-frequency electromagnetic fields, and develop efficient methods for solving the simplified formulations. This is done via the use of an asymptotic method. Of course, we are not the first to seek such asymptotic approaches.

An asymptotic method was first introduced by Rytov \cite{Rytov1940ZhETF}, and also can be found in Jackson's book \cite{Jackson1998book}. Mitzner \cite{Mitzner1967RS} derived a boundary integral equation for this, which enables computing the scattered fields from bodies with thin skin depths with no additional assumptions imposed on the carrier media. In the last two decades, a number of authors proposed the use of boundary element methods for solving problems with eddy currents \cite{Nicolas1988IEEE}-\cite{Phan2020IEEE}. It is typical that in some studies, a quasistatic approximation for the magnetic field is accepted, and only the magnetic field is computed using the zero-order approximation for the internal problem (e.g., \cite{Sun2002IEEE}). In this case, the solution to the scattering problem can be reduced to computing only one scalar quantity, the surface magnetic potential \cite{Sun2002IEEE}. It may be noted that in many studies, the boundary integral equations for eddy current simulations are derived from the general boundary integral equations for Maxwell's equations in terms of vector quantities, which require special surface basis functions, which substantially complicates the method. Combinations of the boundary element method and finite element method, as well as fast multipole accelerations, can also be found \cite{Takahashi2006IEEE}. Mathematical aspects of the problem are considered in several studies (e.g. \cite{Weggler2012MMAS}, \cite{Bonnet2019CMA}).

We develop and demonstrate a new approach to solve low-frequency electromagnetic problems in the case of thin skin depth inside conductors. This approach is free of the limitation, $\Delta \ll \delta $, allowing for much smaller numerical problem sizes. The approach reduces  to solution of a few electrostatic and magnetostatic problems for scalar potentials, and it is much more efficient compared to standard boundary element (method of moments) Maxwell solvers. Moreover, for large problems, the method can be accelerated using the fast multipole method (FMM) for the Laplace equation \cite{Greengard1987JCP}, \cite{Adelman2017IEEE}, which makes it scalable and suitable for parallelization on large computational clusters \cite{Hu2011:SC}. In this paper, a two-term asymptotic solution is obtained, which can be extended to higher orders if needed. The unique feature of  the present method is the use of special surface operators, such as the inverse surface gradient, which enables the reduction of vector problems to scalar problems. The basic formulation is presented in \S\ref{problem}, while its asymptotic solution approach and the corresponding integral equation formulation and validation for the case of a sphere are described in \S\ref{method}.  \S\ref{results} provides the results for a number of more complex cases, including practical cases. \S\ref{conclusions} concludes the paper.

\section{Problem Statement}\label{problem}
\begin{figure}[htb]
	\vspace{-20pt}
	\par
	\begin{center}
		\includegraphics[width=0.9\textwidth, trim=0 1.15in 1.5in 0]{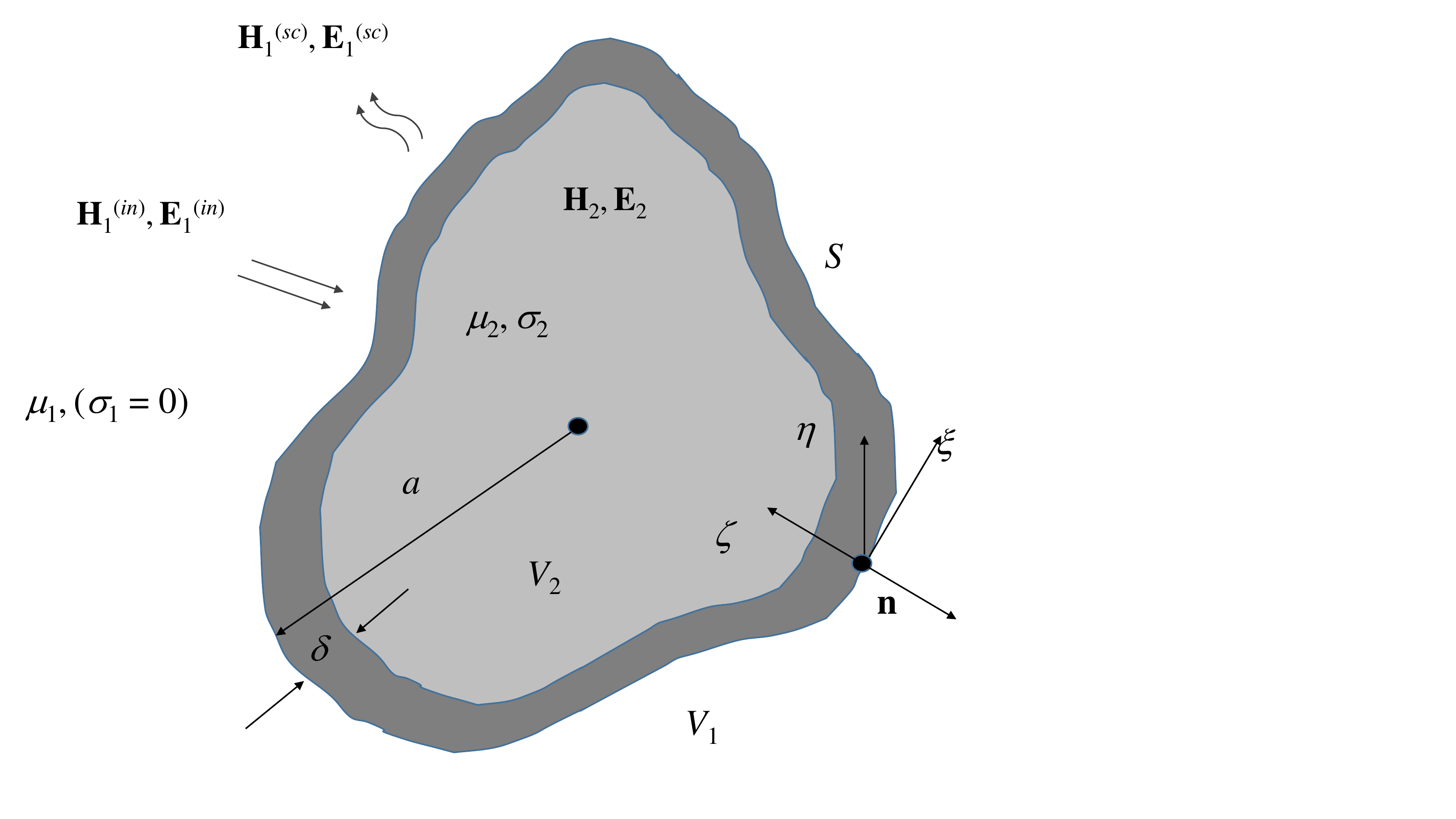}
	\end{center}
	\caption{The problem and notation. An known incident electromagnetic field ${\bf E}^{in}_1,{\bf H}^{in}_1$ from domain $V_1$ with magnetic permeability $\mu_{1}$,  interacts with a scatterer occupying a volume $V_2$ and with electrical conductivity $\sigma_{2}$ and magnetic permeability  $\mu_{2}$, and separated from $V_1$ by the boundary $S$. The interaction causes a field in domain $V_1$ which is a sum of the incident fields, and scattered fields  ${\bf E}^{in}_1+{\bf E}^{sc}_1,{\bf H}^{in}_1+{\bf H}^{sc}_1$; and a field inside the scatterer  ${\bf E}_2,{\bf H}_2$. The characteristic size of domain $V_2$ is $a$, while the parameter $\delta$ is the skin depth, defined later in Eq. (\ref{ps6}). We seek to develop fast approximate integral equation based algorithms to compute ${\bf E}^{sc}_1,{\bf H}^{sc}_1$ and ${\bf E}_2,{\bf H}_2$ for small values of $\delta/a$.}  	\label{Fig0}
\end{figure}
Consider, as shown in Fig. \ref{Fig0}, two low-frequency, time-harmonic electromagnetic fields where the displacement current in Maxwell's equations has been neglected. The first field is contained in the domain, $V_{1}$, possibly extending to infinity, of negligible conductivity, and the second field is contained in the domain, $V_{2}$, of finite conductivity, $\sigma _{2}$, so Maxwell's equations become
\begin{eqnarray}
\nabla \cdot \mathbf{E}_{1} &=&0,\quad \nabla \cdot \mathbf{H}_{1}=0,  \label{ps1} \\
\nabla \times \mathbf{E}_{1} &=&i\omega \mu _{1}\mathbf{H}_{1},\quad \nabla\times \mathbf{H}_{1}=\mathbf{0},  \notag
\end{eqnarray}
\begin{eqnarray}
\nabla \cdot \mathbf{E}_{2} &=&0,\quad \nabla \cdot \mathbf{H}_{2}=0,  \label{ps2} \\
\nabla \times \mathbf{E}_{2} &=&i\omega \mu _{2}\mathbf{H}_{2},\quad \nabla \times \mathbf{H}_{2}=\sigma _{2}\mathbf{E}_{2}.  \notag
\end{eqnarray}%
In these equations, $\mathbf{E}$ and $\mathbf{H}$ are the electric- and magnetic-field vectors, $\mu $ is the magnetic permeability, and $\omega $ is the circular frequency. The field in the domain, $V_{1}$, is composed of the given incident field and the scattered field, 
\begin{equation}
\mathbf{E}_{1}=\mathbf{E}_{1}^{(in)}+\mathbf{E}_{1}^{(sc)},\quad \mathbf{H}_{1}=\mathbf{H}_{1}^{(in)}+\mathbf{H}_{1}^{(sc)},  \label{ps3}
\end{equation}
each of which satisfies equations (\ref{ps1}).  The incident field may also contain some given sources. The problem is to determine the scattered and internal fields, which satisfy the following boundary conditions on the surface, $S$, with the normal, $\mathbf{n}$, separating the domains, $V_{1}$ and $V_{2}$:
\begin{eqnarray}
\mathbf{n}\cdot \mathbf{E}_{2} &=&0,  \label{ps4} \\
\mathbf{n}\times \left( \mathbf{E}_{2}-\mathbf{E}_{1}^{(sc)}\right) &=& \mathbf{n}\times \mathbf{E}_{1}^{(in)},  \notag \\
\mathbf{n}\cdot \left( \mu _{2}\mathbf{H}_{2}-\mu _{1}\mathbf{H}_{1}^{(sc)}\right) &=&\mu _{1}\mathbf{n}\cdot \mathbf{H}_{1}^{(in)},\quad \notag \\
\mathbf{n}\times \left( \mathbf{H}_{2}-\mathbf{H}_{1}^{(sc)}\right) &=&\mathbf{n}\times \mathbf{H}_{1}^{(in)}.  \notag
\end{eqnarray}
The first condition above, that $\mathbf{n}\cdot \mathbf{E}_{2} = 0$, has an intuitive explanation: no eddy currents that are induced in the conductor can leave through the surface. We assume further that the surface, $S$, is closed and smooth, and that the domain, $V_{2}$, is finite. For the infinite domain, $V_{1}$, the scattered field decays at infinity:
\begin{equation}
\lim_{r\rightarrow \infty }\mathbf{E}_{1}^{(sc)}=\mathbf{0},\quad \lim_{r\rightarrow \infty }\mathbf{H}_{1}^{(sc)}=\mathbf{0.}  \label{ps5}
\end{equation}
The obtained solution can easily be modified for the case when $V_{1}$ is finite and $V_{2}$ is infinite, and we will later show how to do this. Furthermore, we characterize the domain, $V_{2}$, using some typical length scale, $a$, which can be related to the domain size or to the rate of change of $\mathbf{n}$ (e.g., reciprocal curvature). The notion of the length scale is actually only needed to ensure that the skin depth in the conductor is small compared to the conductor size, 
\begin{equation}
\delta =\left( \frac{2}{\sigma _{2}\mu _{2}\omega }\right) ^{1/2}\ll a.\label{ps6}
\end{equation}
Note also that such a constraint can hold in the case when $\sigma _{2}$ and $\mu _{2}$ are spatially varying. Such a situation can occur when the conductor is composed of several materials with different properties or there are several different scatterers in the domain, as will arise in some of our application problems.
\section{Method}\label{method}
\subsection{Boundary Integral Formulation for External Problem}
First, consider the external problem in the domain, $V_{1}$. It is clear here that we are in the magnetostatic regime:
\begin{equation}
\mathbf{H}_{1}^{(sc)}=-\nabla \Psi _{1}^{(sc)},\quad \nabla ^{2}\Psi _{1}^{(sc)}=0.  \label{e1}
\end{equation}%
The electric field, however, can not be written this way because the curl of $\mathbf{E}_{1}^{(sc)}$ is not zero. Instead, by the general Helmholtz decomposition, this field can be represented as
\begin{equation}
\mathbf{E}_{1}^{(sc)}=-\nabla \Phi _{1}^{(sc)}+i\omega \mu _{1}\nabla \times \mathbf{C}_{1}^{(sc)}\mathbf{,\quad }\nabla ^{2}\Phi _{1}^{(sc)}=0, \label{e2}
\end{equation}
where $\mathbf{C}_{1}^{(sc)}$ is some vector field. Substituting Eq.\ (\ref{e2}) into Eq.\ (\ref{ps1}), we see that $\mathbf{C}_{1}^{(sc)}$ satisfies the equation,
\begin{equation}
\nabla \times \nabla \times \mathbf{C}_{1}^{(sc)}=\mathbf{H}_{1}^{(sc)}=-\nabla \Psi _{1}^{(sc)}.  \label{e3}
\end{equation}
Denoting 
\begin{equation}
\phi _{1}^{(sc)}=\nabla \cdot \mathbf{C}_{1}^{(sc)}+\Psi _{1}^{(sc)}, \label{e4} \end{equation}
and using the identity, $\nabla ^{2}=\nabla \left( \nabla \cdot \right) -\nabla \times \nabla \times $, we obtain 
\begin{equation}
\nabla ^{2}\mathbf{C}_{1}^{(sc)}=\nabla \phi _{1}^{(sc)}.  \label{e5}
\end{equation}
As any particular solution of this equation can be used, we use
\begin{equation}
\mathbf{C}_{1}^{(sc)}\left( \mathbf{r}\right) =-\int_{V_{1}}G\left( \mathbf{r},\mathbf{r}^{\prime }\right) \nabla _{\mathbf{r}^{\prime }}\phi_{1}^{(sc)}dV\left( \mathbf{r}^{\prime }\right) ,  \label{e6}
\end{equation}
where $G\left( \mathbf{r},\mathbf{r}^{\prime }\right) $ is the free-space Green's function for the Laplace equation,
\begin{equation}
G\left( \mathbf{r},\mathbf{r}^{\prime }\right) =\frac{1}{4\pi \left| \mathbf{r}-\mathbf{r}^{\prime }\right| },\quad \nabla_{\mathbf{r}^{\prime}}^{2}G \left(\mathbf{r},\mathbf{r}^{\prime }\right) =-\delta \left( \mathbf{r},\mathbf{r}^{\prime }\right) ,  \label{e7}
\end{equation}
and $\delta \left( \mathbf{r},\mathbf{r}^{\prime }\right) $ is the delta function in three-dimensional space. Taking the divergence of both sides of Eq.\ (\ref{e6}) and using Eq.\ (\ref{e4}), we obtain an integral equation for the unknown scalar function, $\phi _{1}^{(sc)}$:
\begin{equation}
\phi _{1}^{(sc)}-\Psi _{1}^{(sc)}=-\nabla _{\mathbf{r}}\cdot \int_{V_{1}}G\left( \mathbf{r},\mathbf{r}^{\prime }\right) \nabla _{\mathbf{r}^{\prime }}\phi _{1}^{(sc)}dV\left( \mathbf{r}^{\prime }\right) .
\label{e8}
\end{equation}
The right-hand side of this equation can be transformed using Green's identity as follows:
\begin{eqnarray}
&&-\nabla _{\mathbf{r}}\cdot \int_{V_{1}}G\left( \mathbf{r},\mathbf{r}^{\prime }\right) \nabla _{\mathbf{r}^{\prime }}\phi _{1}^{(sc)}dV\left(\mathbf{r}^{\prime }\right)  \label{e9} \\
&=&-\int_{V_{1}}\nabla _{\mathbf{r}}G\left( \mathbf{r},\mathbf{r}^{\prime}\right) \cdot \nabla _{\mathbf{r}^{\prime }}\phi _{1}^{(sc)}dV\left(\mathbf{r}^{\prime }\right) \notag \\
&=&\int_{V_{1}}\nabla _{\mathbf{r}^{\prime}}G\left( \mathbf{r},\mathbf{r}^{\prime }\right) \cdot \nabla _{\mathbf{r}^{\prime }}\phi _{1}^{(sc)}dV\left( \mathbf{r}^{\prime }\right)  \notag \\
&=&\int_{V_{1}}\left\{ \nabla _{\mathbf{r}^{\prime }}G\left( \mathbf{r},\mathbf{r}^{\prime }\right) \cdot \nabla _{\mathbf{r}^{\prime }}\phi _{1}^{(sc)}+\left[ \nabla _{\mathbf{r}^{\prime }}^{2}G\left( \mathbf{r}, \mathbf{r}^{\prime }\right) +\delta \left( \mathbf{r},\mathbf{r}^{\prime}\right) \right] \phi _{1}^{(sc)}\left( \mathbf{r}^{\prime }\right) \right\} dV\left( \mathbf{r}^{\prime }\right)  \notag \\
&=&-\int_{S}\phi _{1}^{(sc)}\left( \mathbf{r}^{\prime }\right) \frac{\partial G\left( \mathbf{r},\mathbf{r}^{\prime }\right) }{\partial n\left(\mathbf{r}^{\prime }\right) }dS\left( \mathbf{r}^{\prime }\right) +\alpha \left( \mathbf{r}\right) \phi _{1}^{(sc)}\left( \mathbf{r}\right) ,  \notag
\end{eqnarray}
where the normal is directed from $V_{2}$ to $V_{1}$ (which causes a negative sign to appear in front of the surface integral in the last expression), $\alpha \left(\mathbf{r}\right) =1$ for points internal to $V_{1}$ (i.e., $\mathbf{r}\in V_{1}$), and $\alpha \left( \mathbf{r}\right) =1/2$ for points on $S$ (i.e., $\mathbf{r}\in S).$ In the latter case, the surface integral is singular and should be treated in terms of its principal value ($p.v.$). We now introduce the following notation for single- and double-layer potentials:%
\begin{eqnarray}
L\left[ s\right] \left( \mathbf{r}\right) &=&\int_{S}s\left( \mathbf{r}^{\prime }\right) G\left( \mathbf{r},\mathbf{r}^{\prime }\right) dS\left(\mathbf{r}^{\prime }\right) ,  \label{e10} \\
M\left[ s\right] \left( \mathbf{r}\right) &=&p.v.\int_{S}s\left( \mathbf{r}^{\prime }\right) \frac{\partial G\left( \mathbf{r},\mathbf{r}^{\prime }\right) }{\partial n\left( \mathbf{r}^{\prime }\right) }dS\left( \mathbf{r}^{\prime }\right) .  \notag
\end{eqnarray}%
This shows that the unknown variable, $\phi _{1}^{(sc)}$, on the surface can be found by solving the boundary integral equation,
\begin{equation}
M\left[ \phi _{1}^{(sc)}\right] \left( \mathbf{r}\right) +\frac{1}{2}\phi
_{1}^{(sc)}\left( \mathbf{r}\right) =\Psi _{1}^{(sc)}\left( \mathbf{r}%
\right) ,\quad \mathbf{r}\in S\mathbf{.}  \label{e11}
\end{equation}
Note that $\nabla \times \mathbf{C}_{1}^{(sc)}$ can then be expressed via the
scalar, $\phi _{1}^{(sc)}$. Indeed, using Eq.\ (\ref{e6}), we have%
\begin{eqnarray}
\mathbf{C}_{1}^{(sc)}\left( \mathbf{r}\right) &=&-\int_{V_{1}}G\left( 
\mathbf{r},\mathbf{r}^{\prime }\right) \nabla _{\mathbf{r}^{\prime }}\phi
_{1}^{(sc)}dV\left( \mathbf{r}^{\prime }\right)  \label{e12} \\
&=&-\int_{V_{1}}\nabla _{\mathbf{r}^{\prime }}\left[ G\left( \mathbf{r},%
\mathbf{r}^{\prime }\right) \phi _{1}^{(sc)}\right] dV\left( \mathbf{r}%
^{\prime }\right) +\int_{V_{1}}\phi _{1}^{(sc)}\left( \mathbf{r}^{\prime
}\right) \nabla _{\mathbf{r}^{\prime }}G\left( \mathbf{r},\mathbf{r}^{\prime
}\right) dV\left( \mathbf{r}^{\prime }\right)  \notag \\
&=&\int_{S}G\left( \mathbf{r},\mathbf{r}^{\prime }\right) \phi _{1}^{(sc)}%
\mathbf{n}\left( \mathbf{r}^{\prime }\right) dS\left( \mathbf{r}^{\prime
}\right) -\int_{V_{1}}\phi _{1}^{(sc)}\left( \mathbf{r}^{\prime }\right)
\nabla _{\mathbf{r}}G\left( \mathbf{r},\mathbf{r}^{\prime }\right) dV\left( 
\mathbf{r}^{\prime }\right)  \notag \\
&=&L\left[ \mathbf{n}\phi _{1}^{(sc)}\right] -\nabla \int_{V_{1}}\phi
_{1}^{(sc)}\left( \mathbf{r}^{\prime }\right) G\left( \mathbf{r},\mathbf{r}%
^{\prime }\right) dV\left( \mathbf{r}^{\prime }\right) .  \notag
\end{eqnarray}%
Taking the curl and plugging the result into Eq.\ (\ref{e2}), we obtain
\begin{equation}
\mathbf{E}_{1}^{(sc)}=-\nabla \Phi _{1}^{(sc)}+i\omega \mu _{1}\nabla \times
L\left[ \mathbf{n}\phi _{1}^{(sc)}\right] \mathbf{,\quad }\nabla ^{2}\Phi
_{1}^{(sc)}=0.  \label{e14}
\end{equation}%
This equation shows that only the boundary value of the scalar function, $\phi _{1}^{(sc)}$, is
needed to compute the rotational part of $\mathbf{E}_{1}^{(sc)}$, and Eq.\ (%
\ref{e11}) provides a method to determine this function using the magnetic
potential.
Note that, for the Laplace equation, the operator, $M+\left(1/2\right)I$, in Eq.\ (\ref{e11}) is
degenerate. However, this problem can be solved by applying the additional
condition that the average of $\phi _{1}^{(sc)}$ over the surface is
zero (or in the case of several objects, the average over each object surface
is zero). In fact, any constant added to $\phi _{1}^{(sc)}$ does
not affect the value of $\nabla \times L\left[ \mathbf{n}\phi _{1}^{(sc)(1)}%
\right] $.

\subsection{Asymptotic Expansions for Thin Skin Depth}

Now, consider the internal problem in the domain, $V_{2}$. The magnetic field here satisfies the
vector Helmholtz equation,%
\begin{equation}
\nabla ^{2}\mathbf{H}_{2}+\frac{2i}{\delta ^{2}}\mathbf{H}_{2}=\mathbf{0}.
\label{ai1}
\end{equation}%
At small $\delta $, the field will be nonzero only in some vicinity of the surface, so we separate the $\nabla$ operator into parts related to differentiation along and normal to the surface:%
\begin{equation}
\nabla =\nabla _{s}+\mathbf{n}\frac{\partial }{\partial n}.  \label{ai2}
\end{equation}%
The Laplacian can then be written as%
\begin{eqnarray}
\nabla ^{2} &=&\nabla \cdot \left( \nabla _{s}+\mathbf{n}\frac{\partial }{%
\partial n}\right)   \label{ai3} \\
&=&\nabla _{s}^{2}+\nabla \cdot \left( \mathbf{n}\frac{%
\partial }{\partial n}\right) \notag \\
&=&\nabla _{s}^{2}+\left( \nabla \cdot \mathbf{n}\right) \frac{\partial }{%
\partial n}+\left( \mathbf{n}\cdot \nabla \right) \frac{\partial }{\partial n%
}  \notag \\
&=&\nabla _{s}^{2}+2\kappa \frac{\partial }{\partial n}+\frac{\partial ^{2}}{%
\partial n^{2}},  \notag
\end{eqnarray}%
where $\nabla _{s}^{2}$ is the surface Laplace-Beltrami operator and
\begin{equation}
\kappa =\frac{1}{2}\nabla \cdot \mathbf{n}  \label{ai4}
\end{equation}
is the mean surface curvature. Let us introduce curvilinear coordinates, $\left( \xi ,\eta ,\zeta \right) $, fitted to the surface so that the surface corresponds to $\zeta =0$ and $\zeta $ is directed opposite to the external normal to the center, i.e., $\partial /\partial \zeta =-\partial /\partial n$. Since the electromagnetic field changes only within the skin depth from the value on the surface to zero, we have
\begin{equation}
\mathbf{E}_{2}\left( \mathbf{r}\right) =\mathbf{E}_{2}\left( \xi ^{\prime},\eta ^{\prime },\zeta ^{\prime }\right) ,\quad \mathbf{H}_{2}\left( 
\mathbf{r}\right) =\mathbf{H}_{2}\left( \xi ^{\prime },\eta ^{\prime },\zeta
^{\prime }\right) ,\quad \xi
^{\prime }=\frac{\xi }{a},\quad \eta ^{\prime }=\frac{\eta }{a},\quad \zeta ^{\prime }=\frac{\zeta }{\delta },
\label{ts3}
\end{equation}%
so we obtain%
\begin{equation}
\frac{\partial ^{2}}{\partial \zeta ^{\prime 2}}\mathbf{H}_{2}+2i\mathbf{H}%
_{2}=2\kappa ^{\prime }\frac{\delta }{a}\frac{\partial \mathbf{H}_{2}}{%
\partial \zeta ^{\prime }}-\frac{\delta ^{2}}{a^{2}}\nabla _{s}^{\prime 2}%
\mathbf{H}_{2},\quad \nabla _{s}^{\prime }=a\nabla _{s},\quad
\kappa ^{\prime }=\kappa a,  \label{ai5}
\end{equation}%
where $\nabla _{s}^{\prime }$ is the dimensionless surface del operator in
the coordinates marked by primes, and $\kappa ^{\prime }$ is the dimensionless
surface curvature, which is assumed to be of the order of unity. Further, we consider expansions over the small parameter, $\delta /a$, of the form,%
\begin{equation}
\mathbf{H}_{2}=\mathbf{H}_{2}^{(0)}+\frac{\delta }{a}\mathbf{H}_{2}^{(1)}+\left(\frac{\delta }{a}\right)^2\mathbf{H}_{2}^{(2)}+\left(\frac{\delta }{a}\right)^3\mathbf{H}_{2}^{(3)}+\ldots,
\label{ai6}
\end{equation}%
and similarly for all other internal and external fields. Substituting Eq.\ (\ref{ai6}) into Eq.\ (\ref{ai5}) and collecting terms of the same power of $\delta /a$, we obtain the following recurrence relations:
\begin{eqnarray}
\frac{\partial ^{2}}{\partial \zeta ^{\prime 2}}\mathbf{H}_{2}^{(0)}+2i \mathbf{H}_{2}^{(0)} &=&\mathbf{0}, \label{recurrence1} \\
\frac{\partial ^{2}}{\partial \zeta ^{\prime 2}}\mathbf{H}_{2}^{(1)}+2i \mathbf{H}_{2}^{(1)} &=&2\kappa ^{\prime }\frac{\partial \mathbf{H}_{2}^{(0)}}{\partial \zeta ^{\prime }}, \notag \\
\frac{\partial ^{2}}{\partial \zeta ^{\prime 2}}\mathbf{H}_{2}^{(j)}+2i\mathbf{H}_{2}^{(j)} &=&2\kappa^{\prime }\frac{\partial \mathbf{H}_{2}^{(j - 1)}}{\partial \zeta ^{\prime }}-\nabla _{s}^{\prime ^2}\mathbf{H}_{2}^{(j - 2)},\quad j=2,3,\ldots \notag
\end{eqnarray}
In this paper, we focus on the first-order approximation, which requires the first two terms.  They are written here again with boundary conditions imposed:
\begin{eqnarray}
\frac{\partial ^{2}}{\partial \zeta ^{\prime 2}}\mathbf{H}_{2}^{(0)}+2i\mathbf{H}_{2}^{(0)} &=&\mathbf{0},\quad \left. \mathbf{H}_{2}^{(0)}\right|_{\zeta ^{\prime }=0}=\mathbf{H}_{2S}^{(0)},\quad \left. \mathbf{H}_{2}^{(0)}\right| _{\zeta ^{\prime }=\infty }=\mathbf{0,}  \label{ai7} \\
\frac{\partial ^{2}}{\partial \zeta ^{\prime 2}}\mathbf{H}_{2}^{(1)}+2i\mathbf{H}_{2}^{(1)} &=&2\kappa ^{\prime }\frac{\partial \mathbf{H}_{2}^{(0)}}{\partial \zeta ^{\prime }},\quad \left. \mathbf{H}_{2}^{(1)}\right|_{\zeta ^{\prime }=0}=\mathbf{H}_{2S}^{(1)},\quad \left. \mathbf{H}_{2}^{(1)}\right| _{\zeta ^{\prime }=\infty }=\mathbf{0,}\quad  \notag
\end{eqnarray}%
These boundary conditions relate the fields to the
conditions on the surface, marked by the subscript, $S$, and to the conditions
far from the surface, where the field should decay to zero. Solutions to
these equations are%
\begin{eqnarray}
\mathbf{H}_{2}^{(0)} &=&\mathbf{H}_{2S}^{(0)}e^{-(1-i)\zeta ^{\prime }},
\label{ai8} \\
\mathbf{H}_{2}^{(1)} &=&\left( \mathbf{H}_{2S}^{(1)}+\kappa ^{\prime }%
\mathbf{H}_{2S}^{(0)}\zeta ^{\prime }\right) e^{-(1-i)\zeta ^{\prime }}. 
\notag
\end{eqnarray}
\subsubsection{Zero-Order Approximation}
The zero-order approximation corresponds to the case of vanishing skin depth, i.e., $\delta = 0$.
This means that the scattered field in the zero-order approximation is equal to the field scattered by a perfect conductor.
Thus, the magnetic field can be found from the solution to the Laplace equation with Neumann boundary conditions:
\begin{equation}
\mathbf{H}_{1}^{(sc)(0)}=-\nabla \Psi _{1}^{(sc)(0)},\quad \nabla ^{2}\Psi
_{1}^{(sc)(0)}=0,\quad \left. \frac{\partial }{\partial n}\Psi
_{1}^{(sc)(0)}\right| _{S}=-\left. \frac{\partial }{\partial n}\Psi
_{1}^{(in)}\right| _{S}.  \label{zo1}
\end{equation}%
This shows that the normal component of the total external field is zero, so from the boundary conditions in Eq.\ (\ref{ps4}), this means that the normal
component of $\mathbf{H}_{2}^{(0)}$ on the surface is also
zero. The tangenetial component can be found from the solution to Eq.\ (\ref%
{zo1}):%
\begin{equation}
\mathbf{H}_{2S}^{(0)}=-\mathbf{n\times n\times H}_{2S}^{(0)}=-\mathbf{%
n\times n\times }\left. \left( \mathbf{H}_{1}^{(in)}+\mathbf{H}%
_{1}^{(sc)(0)}\right) \right| _{S}=-\nabla _{s}\left( \Psi _{1}^{(in)}+\Psi
_{1}^{(sc)(0)}\right) .  \label{zo2}
\end{equation}%
Using this value, we obtain $\mathbf{H}_{2}^{(0)}$ from Eq.\ (\ref%
{ai8}).

As soon as $\Psi _{1}^{(sc)(0)}$ is available, the auxillary function, $\phi
_{1}^{(sc)(0)}$, responsible for the rotational part of the scattered
electric field can be found from the equation,
\begin{equation}
M\left[ \phi _{1}^{(sc)(0)}\right] \left( \mathbf{r}\right) +\frac{1}{2}\phi
_{1}^{(sc)(0)}\left( \mathbf{r}\right) =\Psi _{1}^{(sc)(0)}\left( \mathbf{r}%
\right) ,\quad \mathbf{r}\in S\mathbf{.}  \label{zo3}
\end{equation}%
Since the electric field inside a perfect conductor is zero,%
\begin{equation}
\mathbf{E}_{2}^{(0)}=\mathbf{0},  \label{zo4}
\end{equation}%
the tangential components of the incident and scattered fields on the
surface are simply related, and according to Eq.\ (\ref{e14}), we have%
\begin{eqnarray}
-\mathbf{n}\times \mathbf{n}\times \mathbf{E}_{1}^{(in)(0)} &=&\mathbf{n}%
\times \mathbf{n}\times \mathbf{E}_{1}^{(sc)(0)} \label{zo5} \\
&=&\mathbf{n}\times \mathbf{n}%
\times \left\{ -\nabla \Phi _{1}^{(sc)(0)}+i\omega \mu _{1}\nabla \times L%
\left[ \mathbf{n}\phi _{1}^{(sc)(0)}\right] \right\}  \notag \\
&=&-\mathbf{n}\times \mathbf{n}\times\nabla\Phi _{1}^{(sc)(0)}+i\omega \mu _{1}\mathbf{n}\times \mathbf{n}%
\times \nabla \times L\left[ \mathbf{n}\phi _{1}^{(sc)(0)}\right] \notag \\
&=&\nabla _{s}\Phi _{1}^{(sc)(0)}+i\omega \mu _{1}\mathbf{n}\times \mathbf{n}%
\times \nabla \times L\left[ \mathbf{n}\phi _{1}^{(sc)(0)}\right] ,\quad 
\mathbf{r}\in S.  \notag
\end{eqnarray}%
The problem here is to obtain $\Phi _{1}^{(sc)(0)}$ from its
surface gradient. Such a problem is solvable, and we introduce the inverse surface del operator, $\mathbf{\nabla }_{s}^{-1}$, and propose an algorithm for computing it, later on in this paper.  Using this, we get
\begin{equation}
\Phi _{1}^{(sc)(0)}\left( \mathbf{r}\right) =\nabla _{s}^{-1}\left\{ -%
\mathbf{n}\times \mathbf{n}\times \left\{ \mathbf{E}_{1}^{(in)(0)}+i\omega
\mu _{1}\nabla \times L\left[ \mathbf{n}\phi _{1}^{(sc)(0)}\right] \right\}
\right\} ,\quad \mathbf{r}\in S\mathbf{.}  \label{zo6}
\end{equation}%
This determines the boundary conditions for the Dirichlet problem. After the
Laplace equation is solved with these Dirichlet boundary conditions, we have $\Phi
_{1}^{(sc)(0)}\left( \mathbf{r}\right) $ for any point in $V_{1}$, and can
compute the scattered electric field using Eq.\ (\ref{e14}):%
\begin{equation}
\mathbf{E}_{1}^{(sc)(0)}\left( \mathbf{r}\right) =-\nabla \Phi
_{1}^{(sc)(0)}\left( \mathbf{r}\right) +i\omega \mu _{1}\nabla \times L\left[
\mathbf{n}\phi _{1}^{(sc)(0)}\right] \left( \mathbf{r}\right) ,\quad \mathbf{%
r}\in V_{1}.  \label{zo7}
\end{equation}
\subsubsection{First-Order Approximation}
The first-order approximation can be constructed by considering
Faraday's law, which, using the primed coordinates in Eqs.\ (\ref{ts3}) and (\ref{ai5}), takes the form,%
\begin{equation}
\frac{1}{a}\nabla _{s}^{\prime }\times \mathbf{E}_{2}-\frac{1}{\delta }%
\mathbf{n}\times\frac{\partial \mathbf{E}_{2}}{\partial \zeta ^{\prime }}%
=i\omega \mu _{2}\mathbf{H}_{2}.  \label{fo1}
\end{equation}%
Plugging in the first-order approximation, we get
\begin{equation}
\frac{1}{a}\nabla _{s}^{\prime }\times\left(\mathbf{E}_{2}^{(0)} + \frac{\delta}{a}\mathbf{E}_{2}^{(1)}\right)
-\frac{1}{\delta }%
\mathbf{n}\times\left(\frac{\partial \mathbf{E}_{2}^{(0)}}{\partial \zeta ^{\prime }} + \frac{\delta}{a}\frac{\partial \mathbf{E}_{2}^{(1)}}{\partial \zeta ^{\prime }}\right)%
=i\omega \mu _{2}\left(\mathbf{H}_{2}^{(0)} + \frac{\delta}{a}\mathbf{H}_{2}^{1}\right).  \label{fo1_2}
\end{equation}%
We can divide this equation into two separate ones:
\begin{equation}
\frac{1}{a}\nabla _{s}^{\prime }\times\mathbf{E}_{2}^{(0)}
-\frac{1}{a }%
\mathbf{n}\times\frac{\partial \mathbf{E}_{2}^{(1)}}{\partial \zeta ^{\prime }}%
=i\omega \mu _{2}\mathbf{H}_{2}^{(0)},  \label{fo1_3}
\end{equation}%
\begin{equation}
\frac{\delta}{a^2}\nabla _{s}^{\prime }\times\mathbf{E}_{2}^{(1)}
-\frac{1}{\delta }%
\mathbf{n}\times\frac{\partial \mathbf{E}_{2}^{(0)}}{\partial \zeta ^{\prime }}%
=\frac{i\omega \mu _{2}\delta}{a}\mathbf{H}_{2}^{(1)}.  \label{fo1_4}
\end{equation}
Finding two solutions to these two equations separately will yield a solution to the original equation when combined.

Starting with Eq.\ (\ref{fo1_3}), since $\mathbf{E}_{2}^{(0)}=\mathbf{0}$ and $\mathbf{E}_{2}$ is tangential
on the surface (see Eq.\ (\ref{ps4})), we have 
\begin{eqnarray}
\frac{\partial \mathbf{E}_{2}^{(1)}}{\partial \zeta ^{\prime }}&=&-\mathbf{%
n}\times\mathbf{ n}\times\frac{\partial \mathbf{E}_{2}^{(1)}}{\partial \zeta
^{\prime }} \label{fo2} \\
&=&i\omega \mu _{2}a\left( \mathbf{n\times H}_{2}^{(0)}\right) \notag \\
&=&-\frac{\omega \mu _{2}a}{2}\frac{\partial ^{2}}{\partial \zeta ^{\prime 2}}%
\left( \mathbf{n\times H}_{2}^{(0)}\right).  \notag
\end{eqnarray}%
Using Eq.\ (\ref{ai8}), we get%
\begin{equation}
\mathbf{E}_{2}^{(1)}=\mathbf{E}_{2S}^{(1)}e^{-(1-i)\zeta ^{\prime }},\quad 
\mathbf{E}_{2S}^{(1)}=\frac{1-i}{2}\omega \mu _{2}a\left( \mathbf{n\times H}%
_{2S}^{(0)}\right) .  \label{fo3}
\end{equation}

Now having this result, we can use Eq.\ (\ref{fo1_4}) to determine the
normal component of the internal magnetic field on the surface:%
\begin{equation}
\mathbf{n\cdot H}_{2S}^{(1)}=\frac{1}{i\omega \mu _{2}a}\mathbf{n}\cdot
\nabla _{s}^{\prime }\times \mathbf{E}_{2S}^{(1)}.  \label{fo4}
\end{equation}%
This determines the boundary conditions for the Neumann problem to solve for the
scattered magnetic field via Eq.\ (\ref{ps4}):%
\begin{equation}
\left. \frac{\partial \Psi _{1}^{(sc)(1)}}{\partial n}\right| _{S}=-\left. 
\mathbf{n}\cdot \mathbf{H}_{1}^{(sc)(1)}\right| _{S}=-\frac{\mu _{2}}{\mu
_{1}}\mathbf{n\cdot H}_{2S}^{(1)}=-\frac{1}{i\omega \mu _{1}a}\mathbf{n}%
\cdot \nabla _{s}^{\prime }\times \mathbf{E}_{2S}^{(1)}.  \label{fo5}
\end{equation}%
This then gives us the
tangential component of the internal magnetic field via Eq.\ (\ref{ps4}):
\begin{equation}
-\mathbf{n\times n\times H}_{2S}^{(1)}=-\left. \mathbf{n\times n\times H}%
_{1}^{(sc)(1)}\right| _{S}=-\left. \nabla _{s}\Psi _{1}^{(sc)(1)}\right|
_{S}.  \label{fo6}
\end{equation}%
All together, the internal magnetic field on the surface is
\begin{equation}
\mathbf{H}_{2S}^{(1)}=-\mathbf{n\times n\times H}_{2S}^{(1)}+\mathbf{n}%
\left( \mathbf{n\cdot H}_{2S}^{(1)}\right) .  \label{fo7}
\end{equation}%
According to Eq.\ (\ref{ai8}), we now have the first-order
approximation for the internal problem.

The last thing to determine in the first-order approximation is the
scattered electric field. The steps to get this from the known magnetic
potential and the boundary conditions are similar to those for the zero-order
approximation. First, we determine the auxillary potential,
\begin{equation}
M\left[ \phi _{1}^{(sc)(1)}\right] \left( \mathbf{r}\right) +\frac{1}{2}\phi
_{1}^{(sc)(1)}\left( \mathbf{r}\right) =\Psi _{1}^{(sc)(1)}\left( \mathbf{r}%
\right).  \label{fo8}
\end{equation}
Then we calculate the the surface
electric potential,
\begin{equation}
\Phi _{1}^{(sc)(1)}=\mathbf{\nabla }_{s}^{-1}\left[ -\mathbf{E}_{2S}^{(1)}-%
\mathbf{n}\times \mathbf{n}\times i\omega \mu _{1}\nabla \times L\left[ 
\mathbf{n}\phi _{1}^{(sc)(1)}\right] \right] ,\quad \mathbf{r}\in S.  \label{fo8_1}
\end{equation}
Finally, we solve the Dirichlet problem to determine the scattered field,%
\begin{equation}
\mathbf{E}_{1}^{(sc)(1)}=-\nabla \Phi _{1}^{(sc)(1)}+i\omega \mu _{1}\nabla
\times L\left[ \mathbf{n}\phi _{1}^{(sc)(1)}\right].  \label{fo9}
\end{equation}

\subsection{Analytical Solution for the Sphere}
\begin{figure}[htb]
	\vspace{-20pt}
	\par
	\begin{center}
		\includegraphics[width=0.9\textwidth, trim=0 2.5in 0.5in 0]{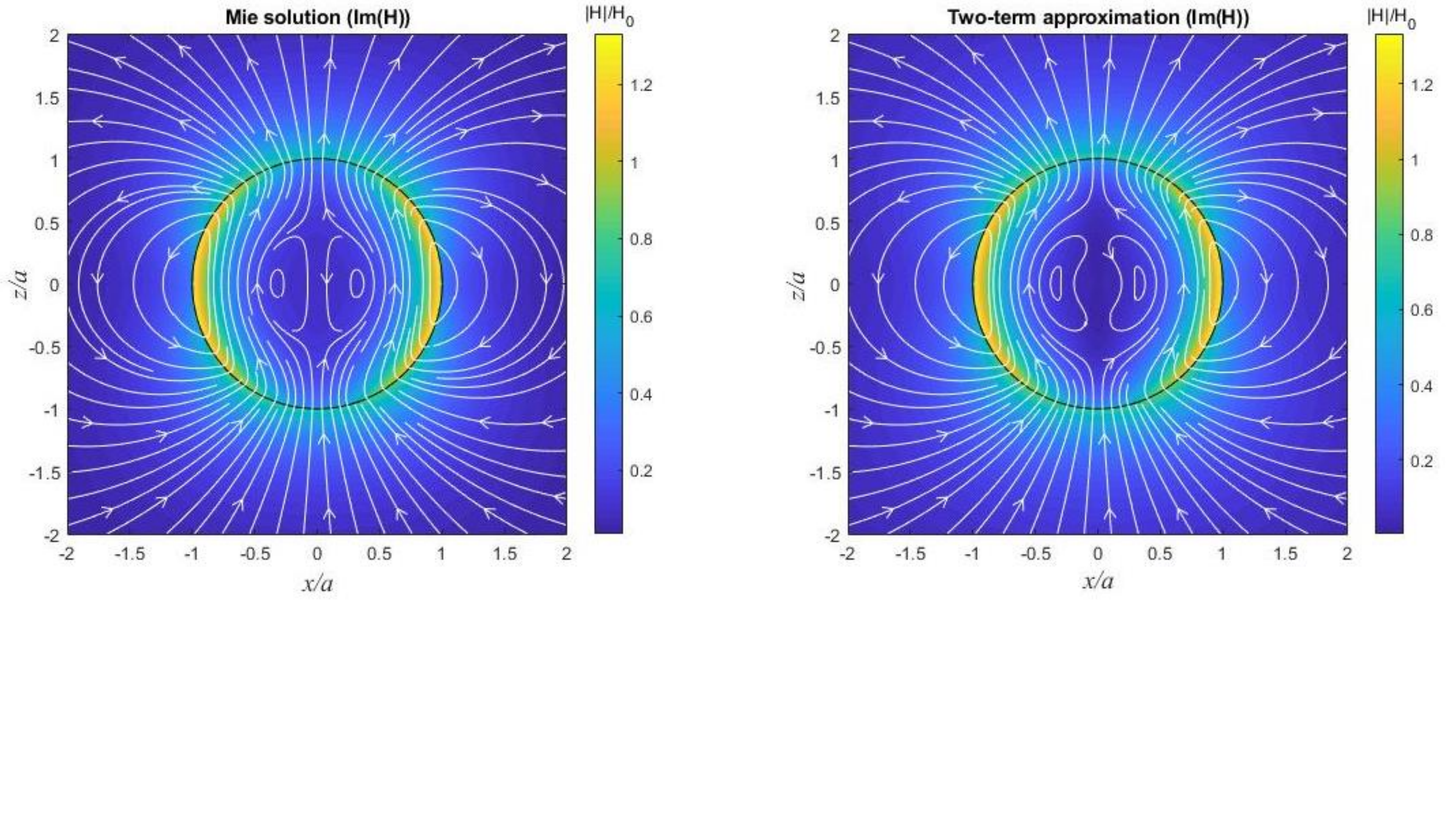}
	\end{center}
	\caption{The Mie solution (left) and the first-order (two-term) approximation from Eq.\ (\ref{a2}) (right) for the
		imaginary part of the magnetic field along the slice, $y=0$. The computations
		were carried out for a copper ball in air at 100 kHz ($a= 1$ mm, $\delta /a=0.2061$). }
	\label{Fig1}
\end{figure}

\begin{figure}[htb]
	\vspace{-20pt}
	\par
	\begin{center}
		\includegraphics[width=0.9\textwidth, trim=0 2.5in 0.5in 0]{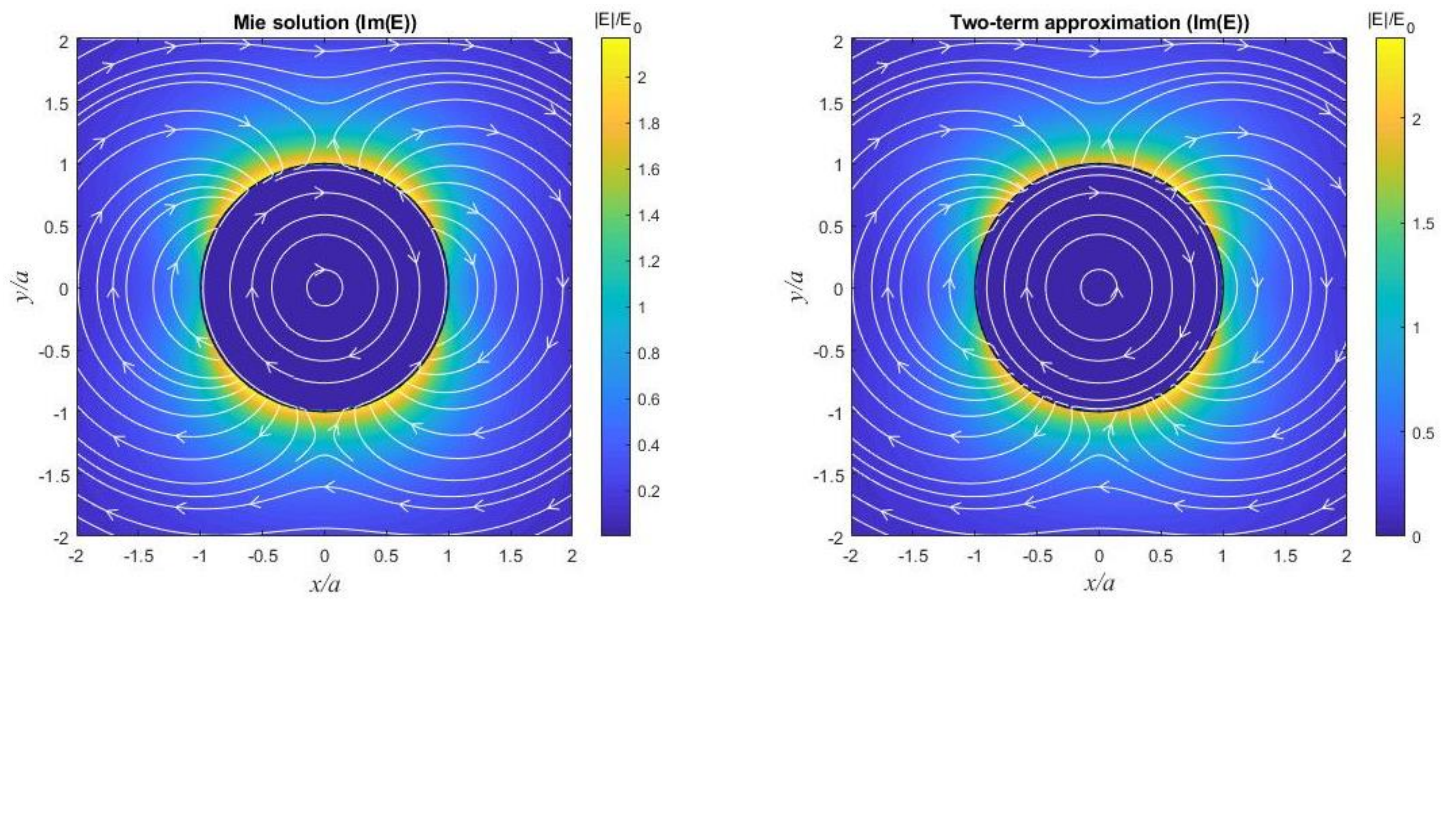}
	\end{center}
	\caption{The Mie solution (left) and the first-order (two-term) approximation from Eq.\ (\ref{a2}) (right) for the
		imaginary part of the electric field along the slice, $z=0$. The computations
		were carried out for a copper ball in air at 100 kHz ($a= 1$ mm, $\delta /a=0.2061$). }
	\label{Fig2}
\end{figure}

It is not difficult to construct an analytical solution, which can be used
for tests. Perhaps the simplest solution is the solution for a sphere of
radius, $a$, illuminated by a plane wave. Assume that the plane wave
propagates in the $x$ direction, and is polarized
in the $y$ direction for the electric field and the $z$ direction for the magnetic
field. In the low-frequency approximation in Eq.\ (\ref{ps1}), we have%
\begin{equation}
\mathbf{E}_{1}^{(in)}=\mathbf{i}_{y}\left( E_{0}+i\omega \mu
_{1}H_{0}x\right) ,\quad \mathbf{H}_{1}^{(in)}=H_{0}\mathbf{i}_{z},\quad
\Psi _{1}^{(in)}=-H_{0}z.  \label{a1}
\end{equation}
Let us introduce spherical coordinates referenced to the center of the sphere:
\begin{equation}
x =r \sin \theta \cos \varphi,  \qquad y =r \sin \theta \sin \varphi, \qquad z =r \cos \theta. \label{a1_5}
\end{equation}
The first-order (two-term) approximation of the problem is
provided below, and its validity can be checked by verifying that it
satisfies the equations and the boundary conditions.
\begin{eqnarray}
\mathbf{E}_{1}^{(sc)(0)} &=&-E_{0}a^{3}\left( \mathbf{i}_{y}\frac{1}{r^{3}}-%
\mathbf{i}_{r}\frac{3y}{r^{4}}\right) \label{a2} \\
&+&i\omega \mu _{1}H_{0}\left[ \frac{a^{3}}{2r^{2}}\mathbf{i}_{r}\times \mathbf{i}_{z}-\frac{a^{5}}{2}\left(\frac{y\mathbf{i}_{x}+x\mathbf{i}_{y}}{r^{5}}-5\frac{xy}{r^{6}}\mathbf{i}
_{r}\right) \right] , \notag \\
\mathbf{E}_{2}^{(0)} &=&0, \notag \\
\mathbf{E}_{1}^{(sc)(1)}&=&-i\omega \mu _{1}c_{1} \frac{H_{0}a^{3}}{2r^{2}}\mathbf{i}_{r}\times \mathbf{i}_{z},  \notag \\
\mathbf{E}_{2}^{(1)}&=&\frac{3\left(1-i\right) }{4}\omega \mu _{2}aH_{0}\left( \mathbf{i}_{r}\times \mathbf{i}_{z}\right) e^{-(1-i)(a-r)/\delta },  \notag \\
\Psi _{1}^{(sc)(0)} &=&-H_{0}\frac{a^{3}}{2r^{3}}z,\quad \mathbf{H}_{1}^{(sc)(0)} =H_{0}\frac{a^{3}}{2r^{3}}\left( \mathbf{i}_{z}-\mathbf{i}_{r}\frac{3z}{r}\right),  \notag \\
\mathbf{H}_{2}^{(0)} &=&\frac{3}{2}H_{0}\left( \mathbf{i}_{z}-\mathbf{i}_{r} \frac{z}{r}\right) e^{-(1-i)(a-r)/\delta }, \notag \\
\Psi_{1}^{(sc)(1)}&=&-c_{1}\Psi _{1}^{(sc)(0)} ,\quad \mathbf{H}_{1}^{(sc)(1)}=-c_{1} \mathbf{H}_{1}^{(sc)(0)},\quad c_{1}=\frac{3\left(1+i\right) \mu _{2}}{2\mu _{1}},  \notag \\
\mathbf{H}_{2}^{(1)}&=&\left(\frac{a-r}{\delta }-\frac{1}{3}c_{1}\right) \mathbf{H}_{2}^{(0)}+\frac{3}{2}H_{0}\left( 1+i\right) \mathbf{i}_{r}\frac{z}{r}e^{-(1-i)(a-r)/\delta }.
\notag
\end{eqnarray}
Apart from this solution, which reflects the approach of this paper, there exists an exact solution of the full problem, which is the Mie solution \cite{Gumerov2007JCP}. To compare
that solution to the present one, the external and internal wavenumbers in that problem should be set to
\begin{equation}
k_{1}=\frac{\omega \mu _{1}H_{0}}{E_{0}},\quad k_{2}=\left( 1+i\right) \sqrt{\frac{\omega \mu _{2}\sigma _{2}}{2}}=\frac{1+i}{\delta }.  \label{a3}
\end{equation}

Figures \ref{Fig1} and \ref{Fig2} compare the Mie solution and the analytical, first-order (two-term) approximation from Eq.\ (\ref{a2}) for the imaginary part of the magnetic and electric fields, respectively. The computations were carried out for a copper ball in air at 100 kHz ($a= 1$ mm, $\delta /a=0.2061$). The lines show the magnetic and electric field lines. It is seen that the two-term solution qualitatively is similar to the Mie solution (the max relative error in the domain shown is roughly 9\%). The colors show the magnitude of the field (the imaginary and real parts are both taken into account), which achieves its maximum in a relatively narrow zone near the boundary. It is also seen that the
magnitude of the electric field is substantially smaller inside the sphere than outside, which is due to $\mathbf{E}_{2}^{(0)}=0$. It is also seen that the internal magnetic field has a non-zero normal component, which is a manifestation of the first-order term, $\mathbf{H}_{2}^{(1)}$ (recall that $\mathbf{H}_{2}^{(0)}$ has only a tangential component). On the other hand, the internal electric field is tangential to the surface, which is also clearly seen in the Mie solution.

\begin{figure}[htb]
\vspace{-20pt}
\par
\begin{center}
\includegraphics[width=0.8\textwidth, trim=0 0.75in 3.5in 0]{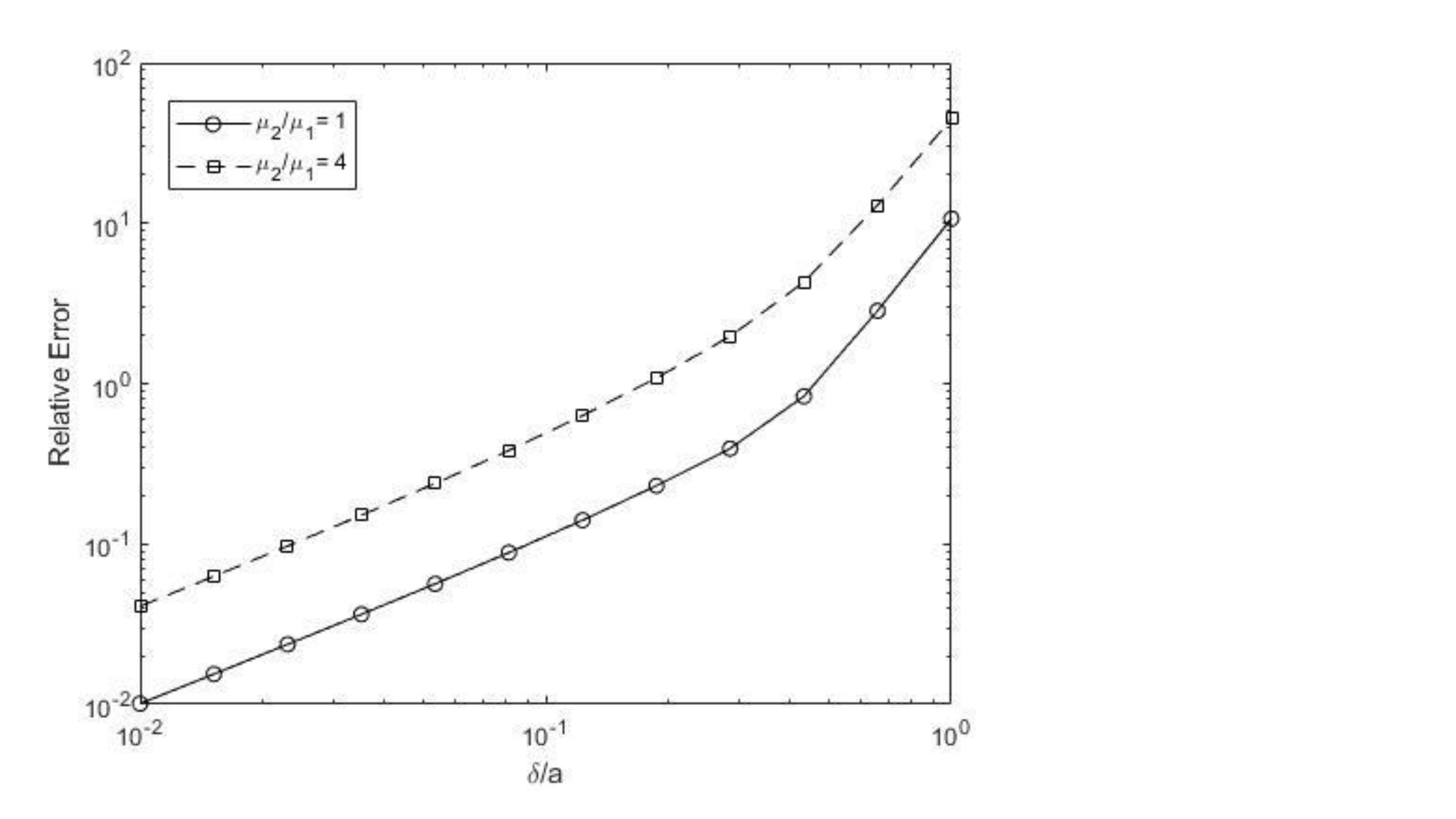}
\end{center}
\caption{The error in $\func{Im}\left\{ H\right\} $ of the asymptotic
solution computed at the surface point \ $z=a$ at different values of $%
\protect\delta /a$ for copper  ($\protect\mu _{2}/\protect\mu _{1}=1$) and
stainless steel ($\protect\mu _{2}/\protect\mu _{1}=4$) balls.}
\label{Fig3}
\end{figure}

Figure \ref{Fig3} illustrates the error of the two-term solution for the imaginary part of the magnetic field computed at the surface point, $z=a$. According to Eq.\ (\ref {a2}), we have 
\begin{equation}
\left. \func{Im}\left\{ \mathbf{H}_{1}^{(sc)}\right\} \right| _{z=a}=\frac{\delta }{a}\left. \func{Im}\left\{ \mathbf{H}_{1}^{(sc)(1)}\right\} \right|
_{z=a}=\frac{3}{2}\frac{\delta }{a}\frac{\mu _{2}}{\mu _{1}}H_0\mathbf{i}_{z}.  \label{a4}
\end{equation}
Values were compared for copper ($\mu_{2}/\mu _{1}=1$, i.e., non-magnetic) and stainless steel ($\mu _{2}/\mu _{1}=4$, i.e., slightly magnetic) balls in the range, $0.01\leqslant \delta /a\leqslant 1$. According to the expansions, the relative error should be $O\left( \delta /a\right) $. The graph shows that this holds. Moreover, for $\mu _{2}/\mu _{1}=1$, the asymptotic constant in $O\left( \delta /a\right) $ is close to one, but for $\mu _{2}/\mu _{1}=4$, it is about $4$. This means that the residual should probably be written as $O\left( \left(\delta/a\right)\left(\mu _{2}/\mu _{1}\right)\right) $. This estimate shows
that for ferromagnetic materials with large $\mu _{2}/\mu _{1}$ ratios, the obtained
solution is applicable only at very small values of $\delta /a$. For
example, the error for carbon steel ($\mu _{2}/\mu _{1}=100$, i.e., very magnetic) was on the
order of one at $\delta /a\approx 10^{-2}$, which is consistent with the
above observation.

\section{Numerical Simulations}\label{results}

To implement the first-order (two-term), low-frequency approximation described
in this paper, we need several numerical tools.
First, solvers for the Laplace equation with Dirichlet and Neumann boundary
conditions
are needed.  A solver for the auxillary equation in Eq.\ (\ref{e11}) is needed
as well. There should also
be routines available for the computation of the curl of the $L$ operator and the
computation of the surface curl, which can be computed using the Stokes theorem.
All of these are available in our previously developed FMM/GPU-accelerated boundary
element method software \cite{Adelman2017IEEE}.
However, the inverse surface gradient is new, and we describe a method for
computing it below.

\subsection{Dirichlet and Neumann Solvers}

Solution of the Laplace equation can be represented in the form of a single
layer potential,%
\begin{equation}
\Phi \left( \mathbf{r}\right) =L\left[ s\right] \left( \mathbf{r}\right) .
\label{de1}
\end{equation}%
So the problem is to determine the single layer density $s$ from the
boundary conditions. For the Dirichlet problem, this reduces to solving of
boundary integral equation%
\begin{equation}
L\left[ s\right] \left( \mathbf{r}\right) =\Phi \left( \mathbf{r}\right)
,\quad \mathbf{r}\in S.  \label{de2}
\end{equation}%
For the Neumann problem the boundary integral equations turns to%
\begin{eqnarray}
L^{\prime }\left[ s\right] \left( \mathbf{r}\right) -\frac{1}{2}s\left( 
\mathbf{r}\right) &=&\frac{\partial }{\partial n}\Phi \left( \mathbf{r}%
\right) ,\quad \mathbf{r}\in S,  \label{de3} \\
L^{\prime }\left[ s\right] \left( \mathbf{r}\right) &=&p.v.\int_{S}s\left( 
\mathbf{r}^{\prime }\right) \frac{\partial G\left( \mathbf{r},\mathbf{r}%
^{\prime }\right) }{\partial n\left( \mathbf{r}\right) }dS\left( \mathbf{r}%
^{\prime }\right) .  \notag
\end{eqnarray}

\subsection{Computation of the Inverse Surface Gradient}

The BEM that we used for the examples in this paper is based on the center panel approximation,
that is, the solution on the boundary is piecewise constant on each panel and the
boundary conditions are enforced at the panel centers.
One step in the low-frequency approximation is to compute the inverse surface gradient, $\phi = \nabla_{s}^{-1}\mathbf{v}$.
In other words, given a surface gradient, $\mathbf{v}$, at the panel centers of a
triangular mesh, we want to determine the potential, $\phi $, on the panels such that $%
\nabla _{s}\phi =\mathbf{v}.$

If $\phi _{1}$ and $\phi _{2}$ are the values of the potential at
neighboring vertices, $\mathbf{x}_{1}$ and $\mathbf{x}_{2}$, of the mesh, then
the directional gradient between them along the edge, $\mathbf{l}$, can be approximated as%
\begin{equation}
\frac{\mathbf{l}}{l}\cdot \mathbf{v}=\frac{\mathbf{l}}{l}\cdot \nabla \phi =%
\frac{d\phi }{dl}\approx \frac{\phi _{2}-\phi _{1}}{l},\quad \mathbf{l=x}%
_{2}-\mathbf{x}_{1},\quad l=\left| \mathbf{l}\right| .  \label{so1}
\end{equation}%
Ideally, in this formula, $\mathbf{v}$ should be evaluated at the point, $\mathbf{x%
}=\left( \mathbf{x}_{1}+\mathbf{x}_{2}\right)/2 $, but since $%
\mathbf{v}$ are available only at the panel centers, we compute $\mathbf{v}$
simply as $\mathbf{v}=\left( \mathbf{v}_{1}+\mathbf{v}_{2}\right) /2
$, where $\mathbf{v}_{1}$ and $\mathbf{v}_{2}$ are the values of $\mathbf{v}$
on the faces sharing $\mathbf{l}$. Hence, each edge produces a linear
equation,%
\begin{equation}
\phi _{2}-\phi _{1}=\frac{1}{2}\left( \mathbf{x}_{2}-\mathbf{x}_{1}\right)
\cdot \left( \mathbf{v}_{1}+\mathbf{v}_{2}\right) .  \label{so2}
\end{equation}%
Since the number of edges in the mesh is larger than the number of vertices, this
forms an overdetermined system, which can be solved using least squares.
However, this problem does not have a unique solution.
Any constant can be added to $\phi$ and its gradient will still equal $\mathbf{v}$.
Thus, the obtained system can be poorly conditioned. For a simply
connective surface, an additional constraint setting the average of $\phi $ over the surface
to zero can be added to the system.
This constraint has an intuitive explanation: if the scattered field is generated by an external field, it should not have any monopole component, meaning $\phi$ should average to zero over the surface.
The equation enforcing this additional constraint is
\begin{equation}
\sum_{j=1}^{N_{v}}w_{j}\phi _{j}=0,\quad w_{j}=\frac{1}{3}\sum_{i}A_{j}^{(i)},
\label{so3}
\end{equation}%
where $w_{j}$ is the weight (surface area) associated with the $j$th vertex, 
$A_{j}^{(i)}$ are the areas of the triangles sharing the $j$th vertex, and $%
N_{v}$ is the total number of vertices. For multi-connective surfaces
(several objects), each object should be supplied by a similar condition, as
the rank of the system is $N_{v}-M$, where $M$ is the number of single
connective surfaces constituting $S$. Note also that technically, it is
simpler to assign some value to some vertex, say $\phi _{1}=0$, find a
solution, and then correct all of the values to a given average.
In any event, as soon as the values at the vertices are determined, the values at
the panel centers can be found using a simple vertex-to-face interpolation:
\begin{equation}
\phi =\frac{1}{3}\left( \phi _{1}+\phi _{2}+\phi _{3}\right) .  \label{so4}
\end{equation}%

The last thing to mention is that this method can naturally be extended to the case
when the BEM uses vertex collocation.  In this case, it becomes even
simpler.

\subsection{Examples}
To support the present development we computed and analysed several cases, some of which are briefly described below. In all examples the incident field was generated in air, which is a dielectric with permittivity $\epsilon_1 =8.85\times 10^{-12}$ F/m and permeability $\mu_1=1.257\times 10^{-6}$ H/m. In the cases illustrated in Fig \ref{Fig4} - Fig\ref{Fig11} the incident electric field is a plane wave of unit intensity (e.g., $E_0 = 1$ V/m), which results in the value $H_0 = \sqrt{\frac{\epsilon_1}{\mu_1}} E_0 = 2.65\times 10^{-3}$ A/m. Also, in these figures the pictures show the internal fields inside the scatterers and the scattered fields outside.

\subsubsection{Sphere}

\begin{figure}[htb]
\vspace{-20pt}
\par
\begin{center}
\includegraphics[width=0.9\textwidth, trim=0 0in 1.5in 0]{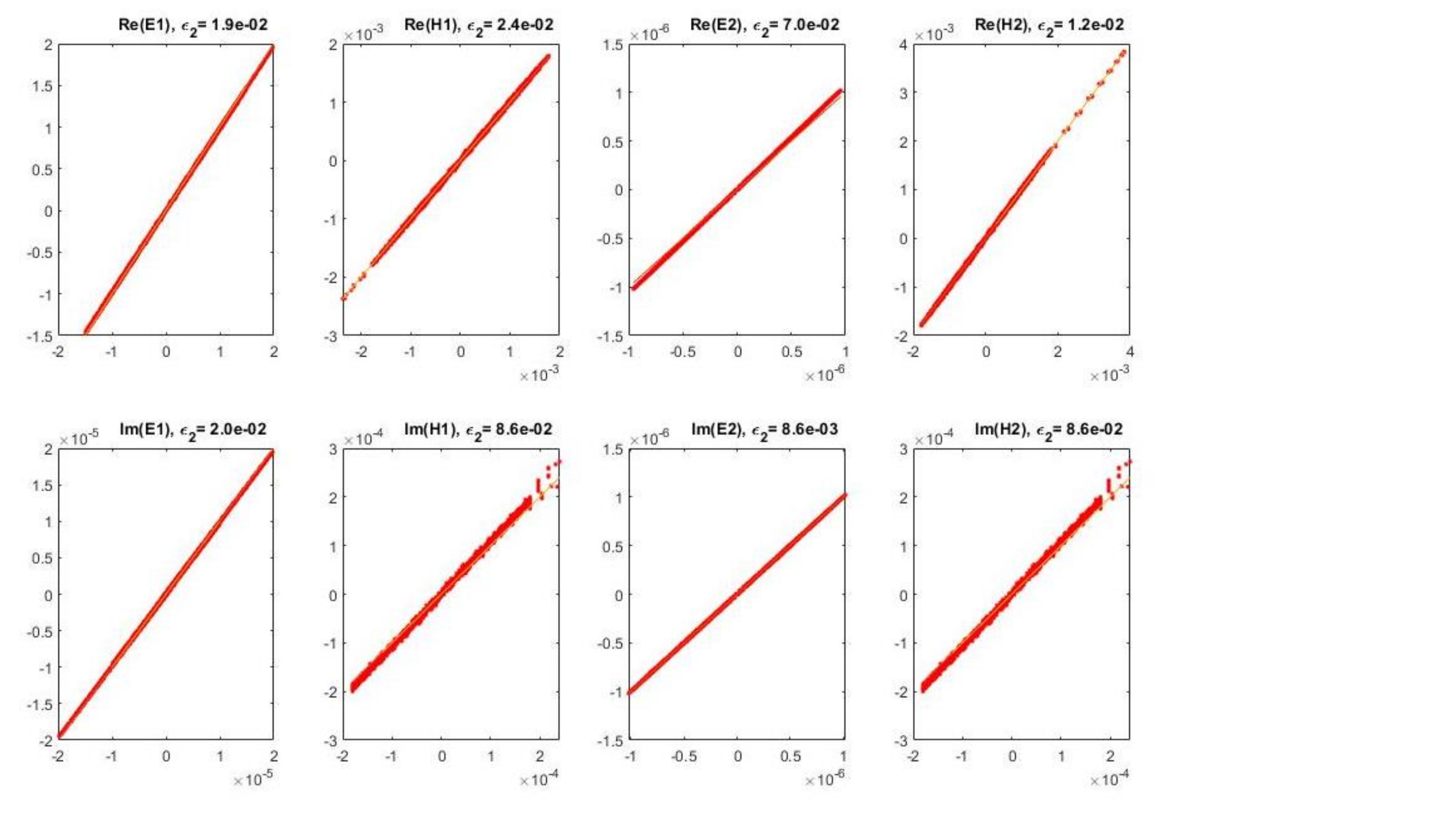}
\end{center}
\caption{The scattered and internal surface fields compared with the exact
Mie solution at the same points (copper, $f=1$ MHz, $a=$1 mm, $\protect%
\delta /a=0.06512$). Ideally these plots should be straight lines described
by $y=x$. }
\label{Fig4}
\end{figure}
First, we computed the benchmark case for the sphere and compared it
to the analytical solution described above and also with the Mie solution.
The obtained results show errors on the order of a few percent for low
discretizations, and the accuracy of the solution increases for higher
discretizations, consistent with
the BEM accuracy described in our recent paper \cite{Adelman2017IEEE}.
In the tests, we used $\delta /a$ in the range, $0.01$--$0.2$ ,and meshes with
$10^{3}$--$10^{6}$ faces. A typical example is shown in Fig. \ref{Fig4},
which plots the scattered and internal surface fields as a function of the
Mie solution at the same points.  In this example, the sphere was made of
copper, had a radius of 1 mm, and the frequency was 1 MHz, so $\delta /a=0.06512$.
Ideally, these plots should be straight lines
described by $y=x$. One can see that the relative RMS error, $\epsilon
_{2}$, of the BEM computations performed on a mesh with 3184 faces is about
2\%. Indeed, the real parts of $\mathbf{E}_{1}^{(sc)}$ and $\mathbf{H}%
_{1}^{(sc)}$ are computed with $O\left( \left( \delta /a\right) ^{2}\right) $
errors, which is about 0.5\%, so the observed errors are due to the BEM
iteslf. On the other hand, the imaginary parts of these fields, as well as
the internal fields, are computed with $O\left( \delta /a\right) $ errors
compared to the same quantites of the Mie solution. So, one can expect
errors on the order of 6-9\% (i.e., the sum of the BEM and the approximation errors
in the worst case). Indeed, such errors are observed for the imaginary part
of the magnetic field and the real part of the electric field, while the
real part of $\mathbf{H}_{2}$ and the imaginary part of $\mathbf{E}_{2}$ are
computed more accurately. This likely indicates that the
second-order approximation should correct $\func{Im}\left( \mathbf{H}%
_{1}\right) $, $\ \func{Im}\left( \mathbf{H}_{2}\right) $, and $\func{Re}%
\left( \mathbf{E}_{2}\right) $, but will not affect $\func{Re}\left( \mathbf{H}%
_{2}\right) $ and $\func{Im}\left( \mathbf{E}_{2}\right) $.  Other
explanations could also be possible.

\subsubsection{Non-Spherical Shapes of Different Topology}

\begin{figure}[htb]
\vspace{-20pt}
\par
\begin{center}
\includegraphics[width=0.9\textwidth, trim=0 1.5in 0.5in 0]{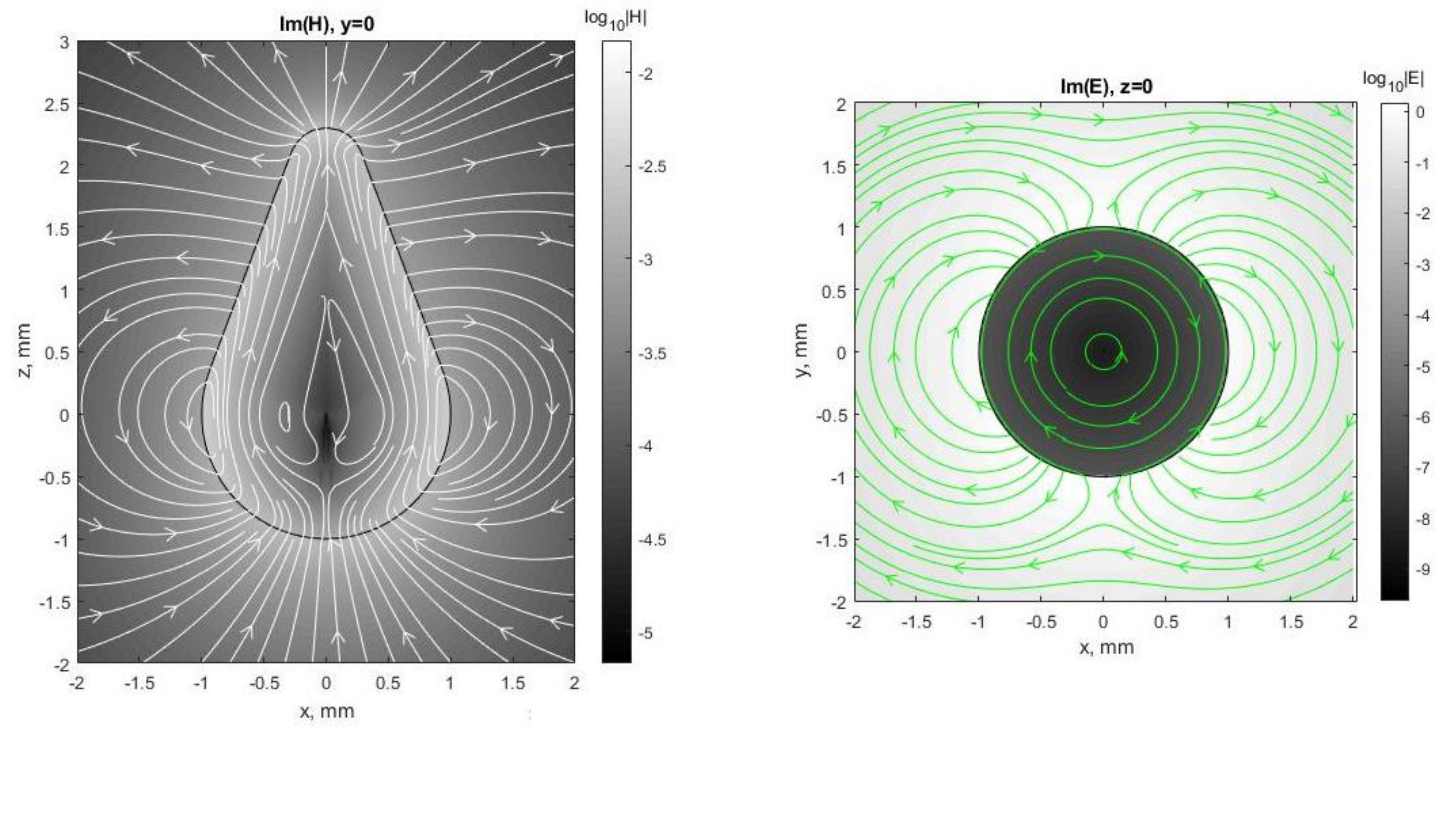}
\end{center}
\caption{The imaginary parts of the scattered and internal fields (copper, $%
f=$ 0.1 MHz, $\protect\delta =0.2061$ mm). The pear shape shown in the
figure was generated as a convex hull of two spheres of radii 1 and 0.3 mm
which centers are separated by the distance 2 mm. }
\label{Fig5}
\end{figure}
\begin{figure}[htb]
\vspace{-10pt}
\par
\begin{center}
\includegraphics[width=0.9\textwidth, trim=0 2.5in 0.5in 0]{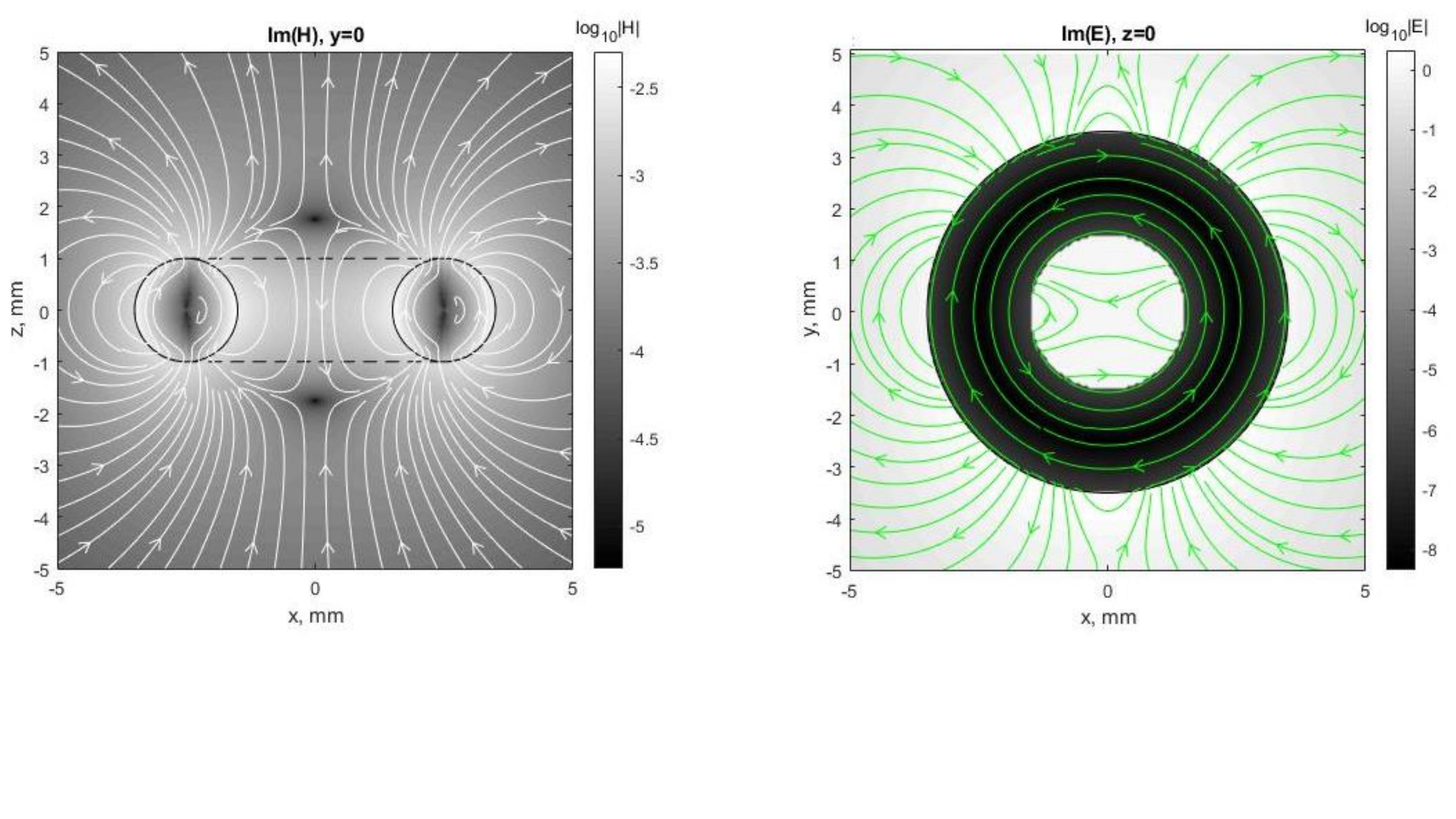}
\end{center}
\caption{The imaginary parts of the scattered and internal fields (copper, $%
f=$ 0.1 MHz, $\protect\delta =0.2061$ mm). The torus larger and small
cross-section radii are 2.5 and 1 mm, respectively. }
\label{Fig6}
\end{figure}

Next, we used simple, non-spherical shapes of different topology,
including a pear-shaped body and a torus, which are bodies of rotation.
Figures \ref{Fig5} and \ref{Fig6} illustrate the imaginary part of the
scattered and internal fields in response to the incident field given by
Eq.\ (\ref{a1}). The pear shape was generated
as the convex hull of two spheres of radii 1 and 0.3 mm with centers
separated by 2 mm. The major and minor radii of the torus are 2.5 and 1 mm,
respectively. The material for both shapes was
copper, and the frequency of the field was 0.1 MHz, so $\delta =0.2061$ mm.
Note that on this and the following figures showing the field lines, there can
be some asymmetry introduced by the plotter (which automatically determines
the position of the field lines to fill the picture more or less evenly). In
these pictures, the shades of gray show the magnitude of the field on a logarithmic
scale (this is also applicable to the figures in the next subsection). The
mesh used contains 6514 faces for the pear shape and
6480 faces for the torus.

\begin{figure}[htb]
\vspace{-20pt}
\par
\begin{center}
\includegraphics[width=0.9\textwidth, trim=0 0.5in 3.5in 0]{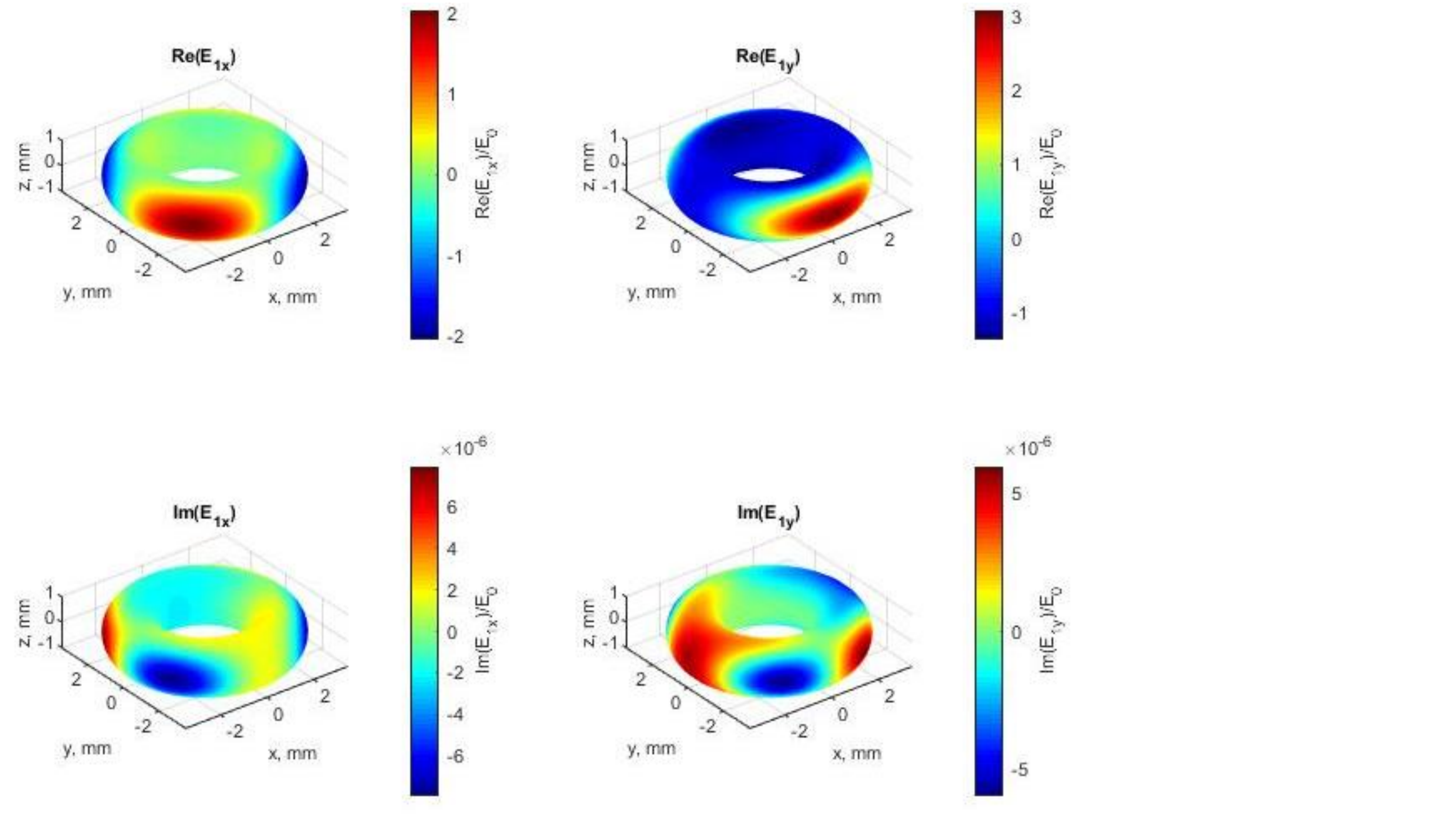}
\end{center}
\caption{The real and imaginary parts of the electric field on the torus
surface. The same case as in Fig. \ref{Fig6}.}
\label{Fig7}
\end{figure}

One peculiarity of the pear shape is that the curvature changes
significantly, which, according to Eq.\ (\ref{ai8}), has an effect on the
internal magnetic field. The internal field is computed well within the skin
depth, while closer to the center of the body, the field lines of the
asymptotic solution can be substanitally distorted. However, since the
magnitude of the field decays exponentially towards the center, such
deviations from the true shapes are not so important. It is seen that while
the magnetic field penetrates the body surface smoothly (at $\mu _{2}=\mu
_{1}$), the magnitude of the electric field inside and outside the conductor
are substantially different. Also, the internal electric field, and so the
electric curent, is tangential to the surface. Figure \ref{Fig7}
shows the real and imaginary parts of the electric field on the surface. It
is remarkable that in the present example, the magnitudes of the real and
imaginary parts differ by a factor of 10$^{5}$. Nevertheless, the present method
handles this situation and produces robust results: the error in
each component is on the order of 1\%.

\subsubsection{Multi-Connective Domains}

\begin{figure}[htb]
\vspace{-20pt}
\par
\begin{center}
\includegraphics[width=0.9\textwidth, trim=0 0.5in 1.5in 0]{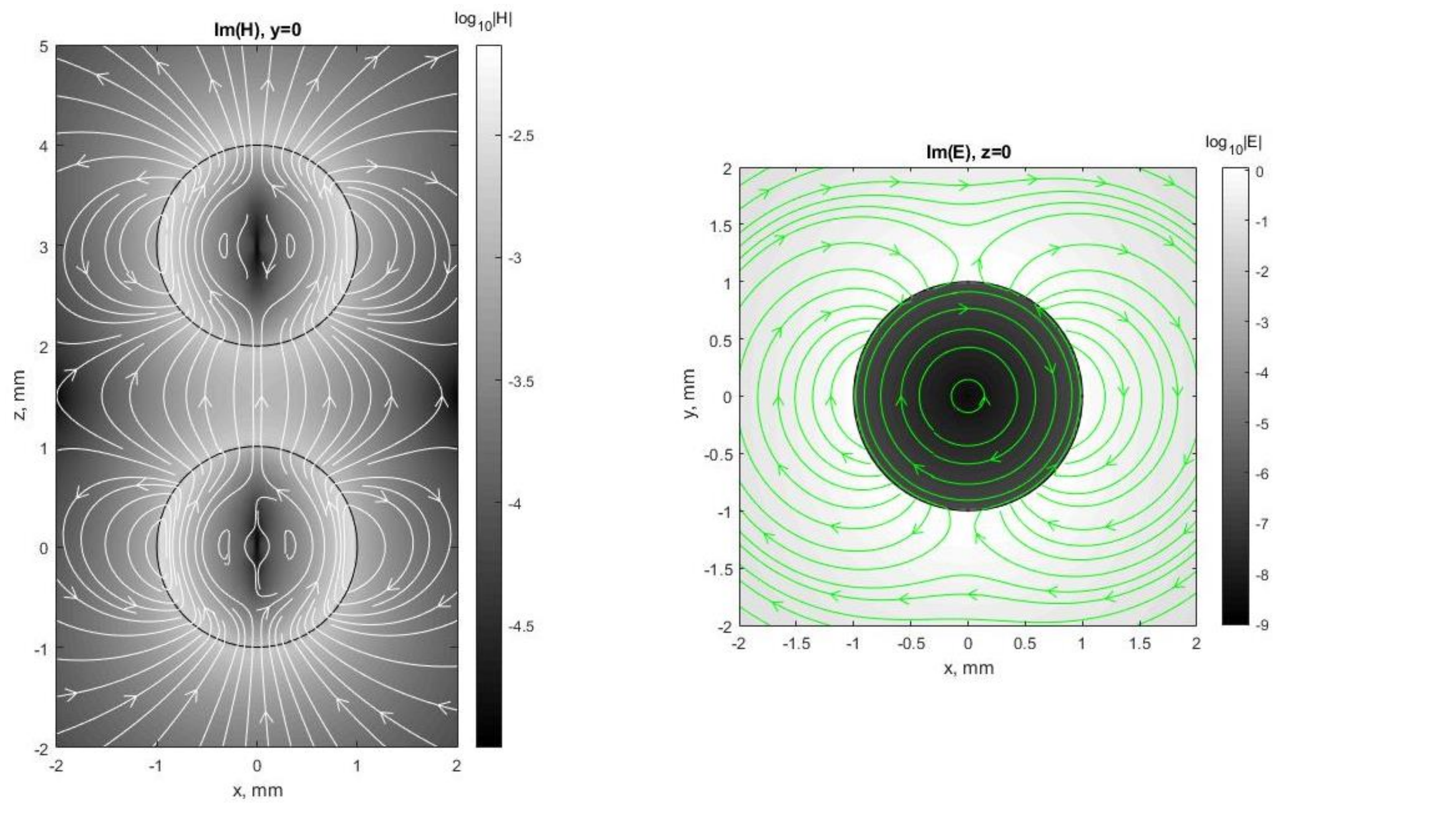}
\end{center}
\caption{The imaginary parts of the scattered and internal fields for two
spheres of radii 1 mm (copper, $f=$ 0.1 MHz, $\protect\delta =0.2061$ mm). }
\label{Fig8}
\end{figure}

\begin{figure}[htb]
\vspace{-10pt}
\par
\begin{center}
\includegraphics[width=0.9\textwidth, trim=0 2.5in 0.5in 0]{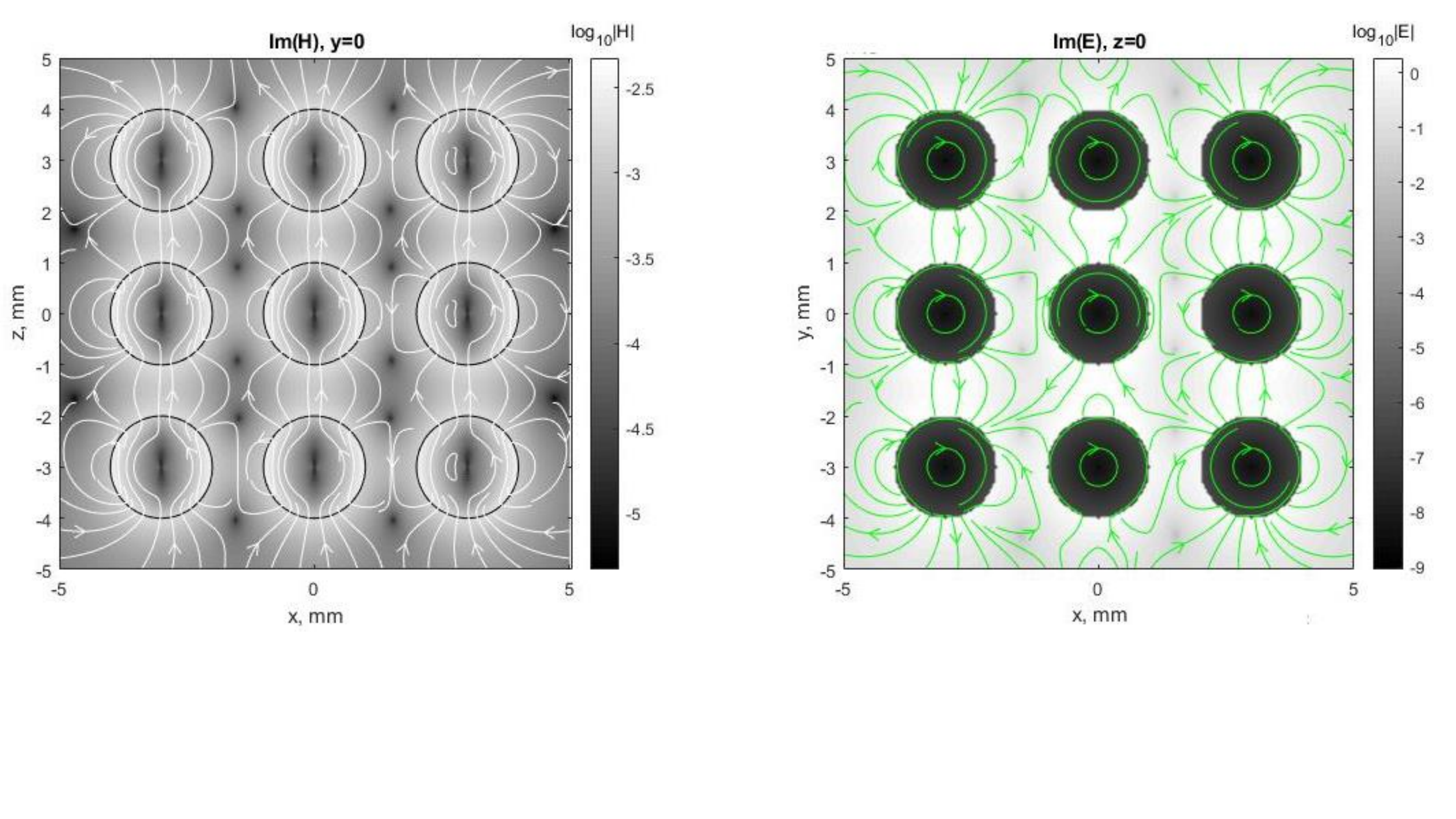}
\end{center}
\caption{The same as in Fig. \ref{Fig8}, but for 27 spheres of equal size $a$
= 1 mm, which centers are located in a cubic grid (min inter-center distance
3 mm).}
\label{Fig9}
\end{figure}

To illustrate that the method works for domains consisting of several
disconnected objects (multiply connective domains), we conducted computations
for several spheres (from 2 to 27). As it was
mentioned above, the modifications of the code here are related to the fact
that the rank deficiency of the matrices, $M+\frac{1}{2}I$, and the inverse
surface gradient is exactly the number of the disconnected objects in the domain, so
additional equations, such as specifying the zero average over each object
should be added. The results obtained show that this approach works well,
and there were no difficulties to compute such cases.

In the cases illustrated in Figs \ref{Fig8} and \ref{Fig9}, the spheres are of the
same radius ($a=$ 1 mm and $\delta /a=0.2061$). There were 3184 faces mesh per sphere.
Note that in the case of 27 spheres, the total number of faces is 85968.
The spheres were made of copper, and the frequency was 1 kHz.
Such cases cannot be normally computed using the BEM on desktop PCs,
but FMM acceleration enables such computations. In all cases, the incident
field is given by Eq.\ (\ref{a1}).

\subsubsection{Real-World Examples}

\begin{figure}[htb]
\vspace{-20pt}
\par
\begin{center}
\includegraphics[width=0.75\textwidth, trim=0 0.75in 3.5in 0]{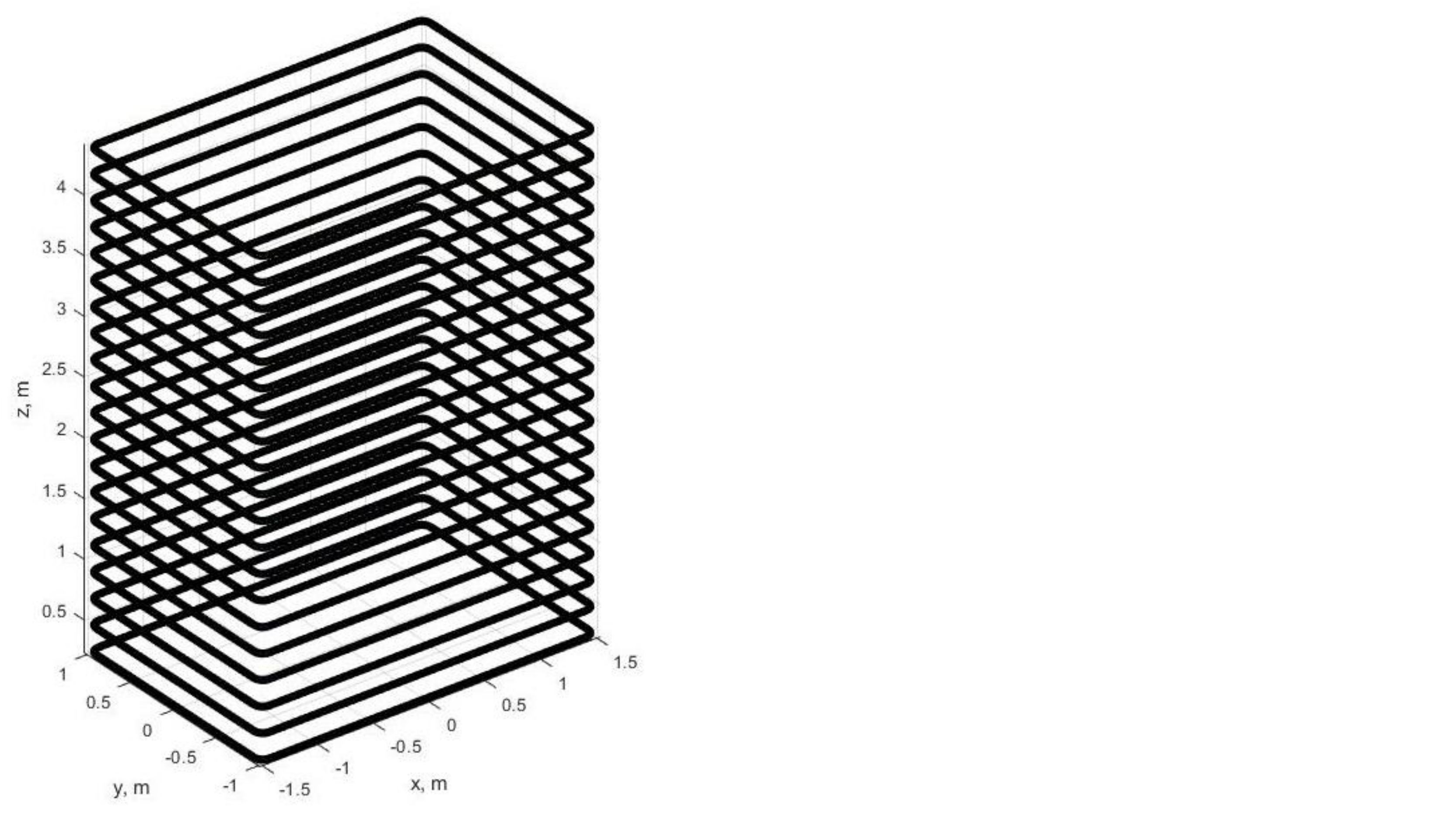}
\end{center}
\caption{The computational mesh for the ARL electric cage (20 loops, 120000
faces, the circular cross-section radius for a rod is 2.54 cm).}
\label{Fig10}
\end{figure}

\begin{figure}[htb]
\vspace{-20pt}
\par
\begin{center}
\includegraphics[width=0.9\textwidth, trim=0 1in 1.5in 0]{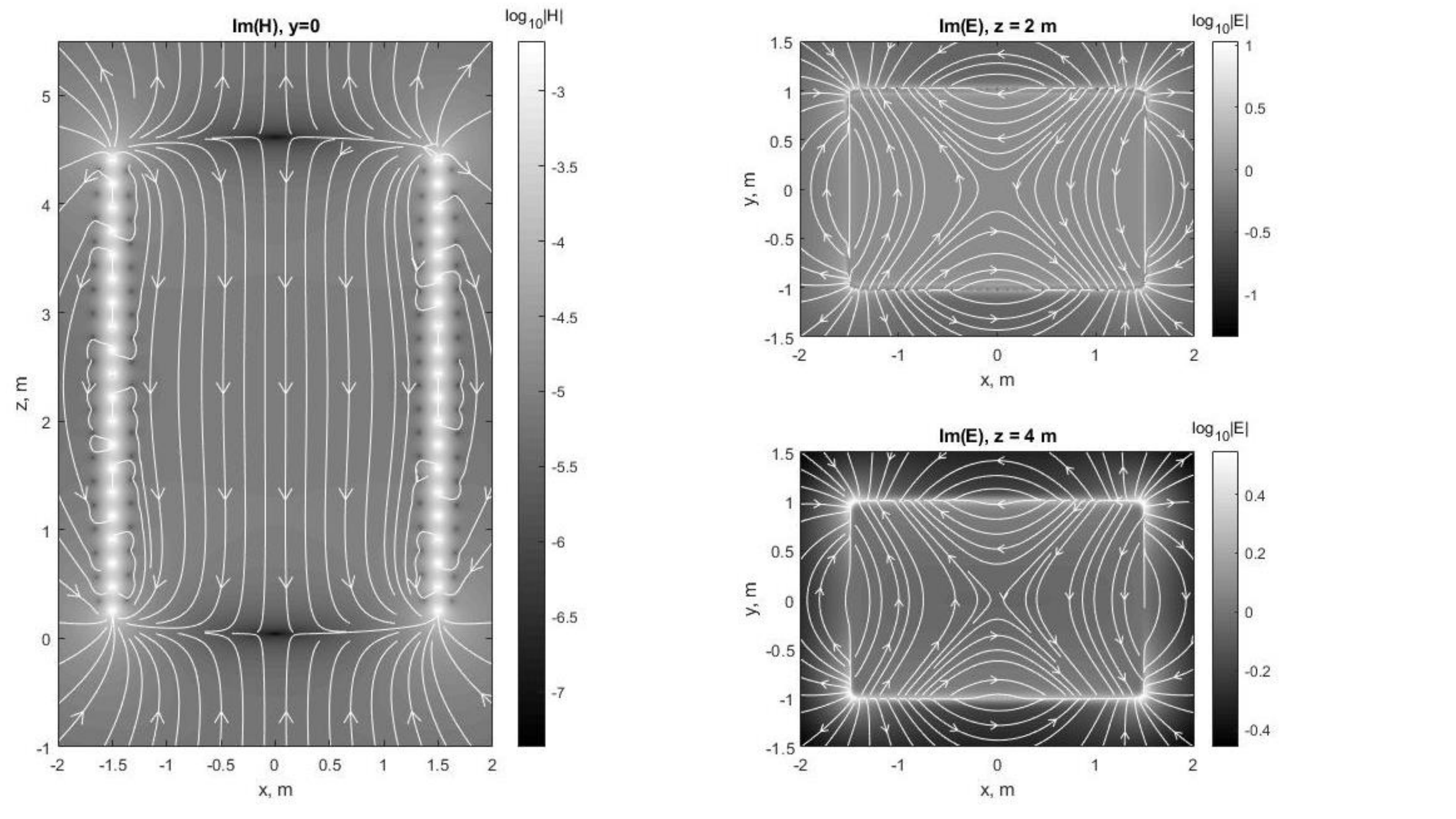}
\end{center}
\caption{The imaginary parts of the scattered fields for the cage mesh shown
in Fig. \ref{Fig10} (aluminum, $f=$ 1 kHz, $\protect\delta =2.6$ mm). }
\label{Fig11}
\end{figure}

Apart from these canonical and simple shapes, we can also use the methods in
this paper to solve real-world problems.
The Army Research Laboratory (ARL) has two facilities for generating
low-frequency electric and magnetic fields for sensor calibration and characterization,
as well as for hardware-in-the-loop experiments.
The electric-field cage, constructed in 2006, generates a uniform, single-axis
electric field to a high degree of accuracy \cite{Hull2006}.
It is composed of two large parallel plates separated by 4.2 m with 20 equally
space ``guard tubes'' between them to control the fringing fields.
The guard tubes are made of aluminum and are 2 in thick (cross-sectional radius
of 2.54 cm).
The mesh
of the electric-field cage,
shown in Fig. \ref{Fig10}, contains 120000 faces (20 guard tubes with 6000 faces
per guard tube).
For testing purposes, we illuminated the electric-field cage with the incident
field in Eq.\ (\ref{a1}).
The magnetic and
electric fields for the cage are shown in Fig. \ref{Fig11}.
Computations were performed at 1 kHz ($\delta =2.6$ mm$)$.
\begin{figure}[htb]
	\vspace{-20pt}
	\par
	\begin{center}
		\includegraphics[width=0.9\textwidth, trim=0 0.5in 2.5in 0]{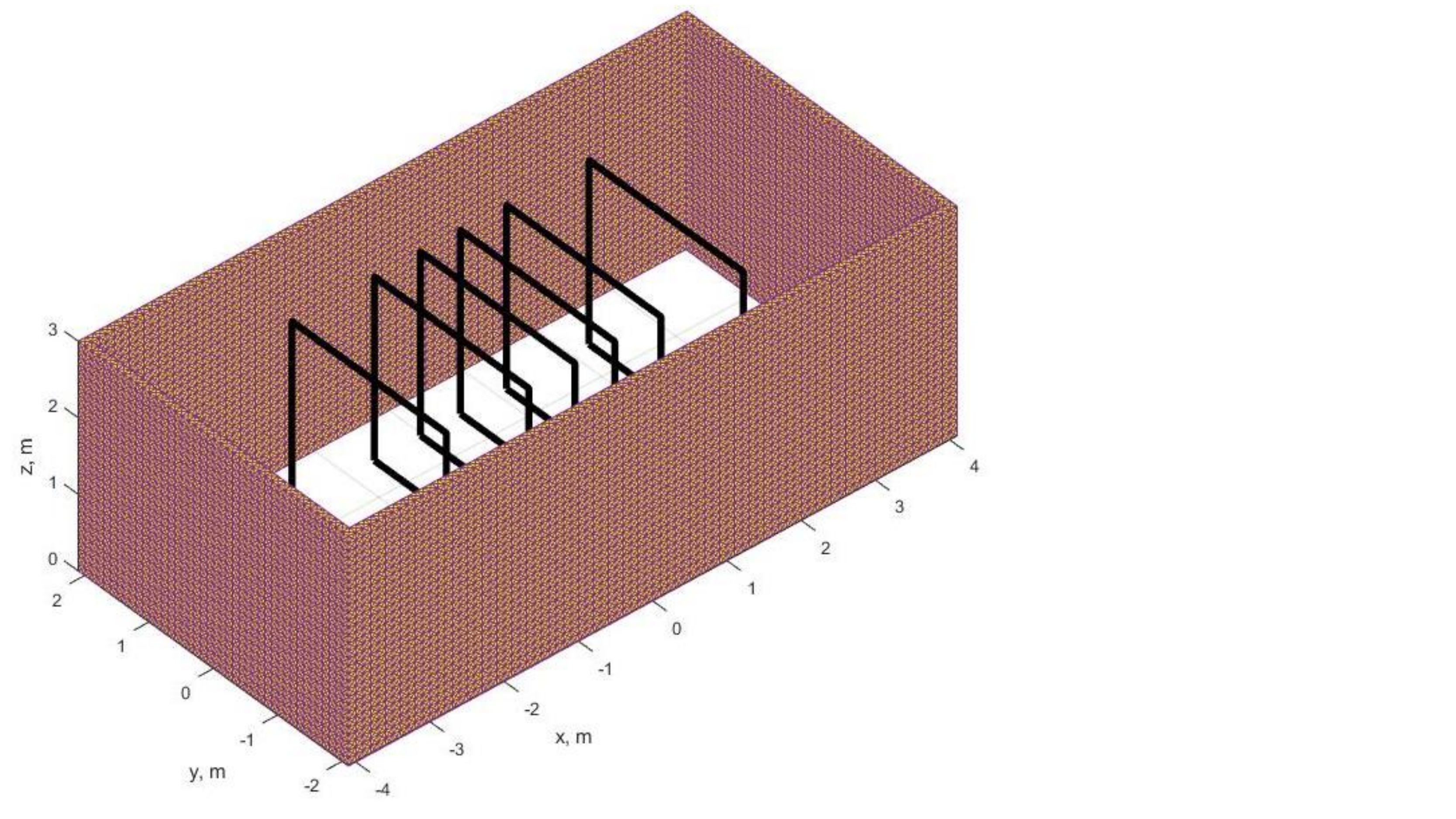}
	\end{center}
	\caption{The computational mesh for the ARL magnetic cage (6
		coils, 60512 faces).}
	\label{Fig7.1}
\end{figure}
\begin{figure}[htb]
	\vspace{-20pt}
	\par
	\begin{center}
		\includegraphics[width=0.9\textwidth, trim=0 0.5in 0.5in 0]{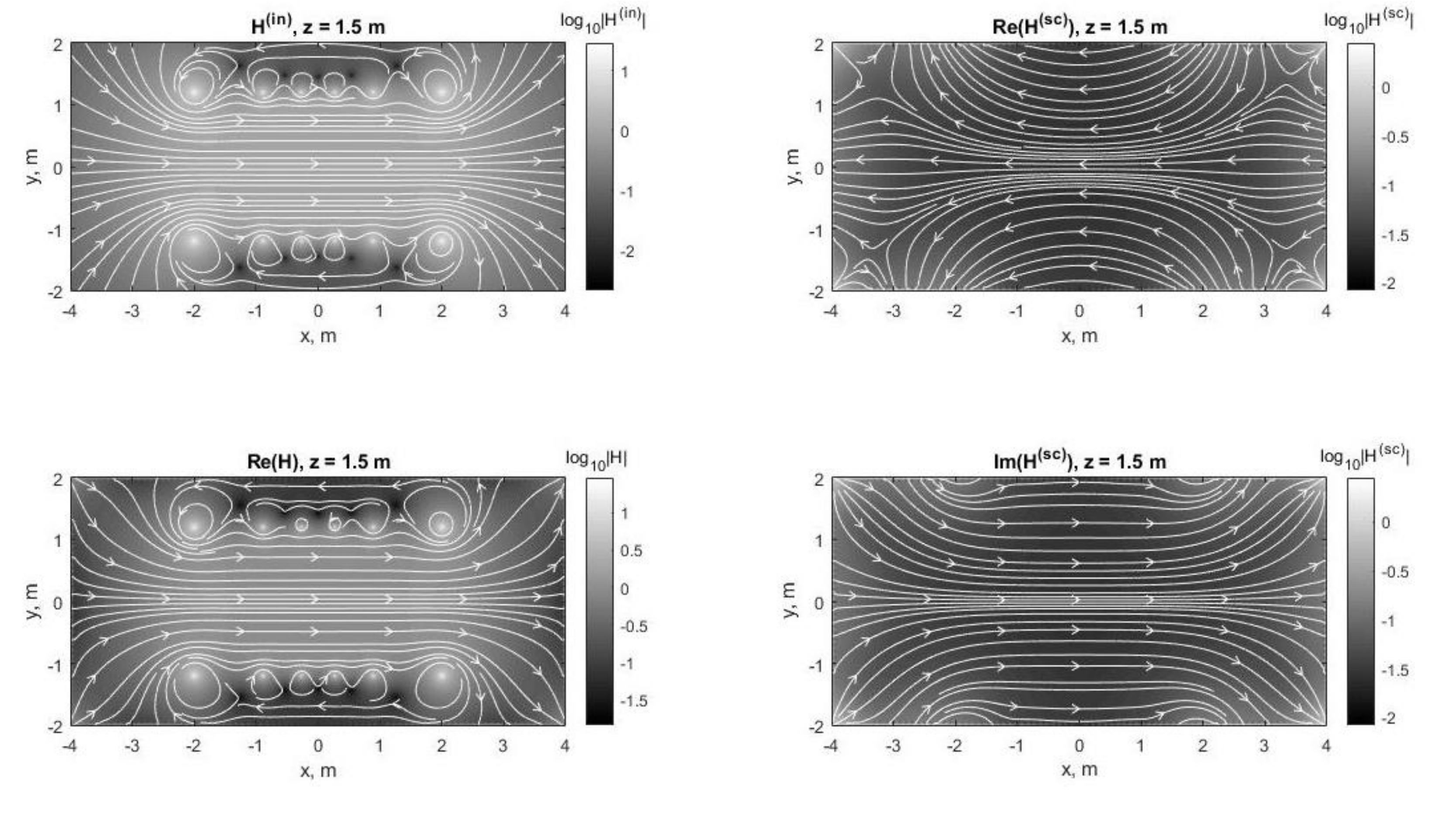}
	\end{center}
	\caption{The incident magnetic field, the total magnetic field, and the real and imaginary parts of the scattered magnetic field for the cage mesh shown in Fig. \ref{Fig7.1} (carbon steel, $f=$ 60 Hz, currents in the coils are 2.0520 A, 0.8250 A, 0.6006 A, 0.6006 A, 0.8250 A, 2.0520 A (as $x$ increases), clockwise orientation for a viewer located at $x=-4$).  }
	\label{Fig7.2}
\end{figure}

The magnetic-field cage, constructed in 2017, generates a uniform, three-axis magnetic field, and is used for similar purposes. It is a coil system, similar to a Helmholtz or Merritt coil, with six coils in the $x$ direction and two coils in the $y$ and $z$ directions  (carbon steel, $\mu _{2}/\mu _{1}=100,$ $\sigma_{2}=6.99\cdot 10^{6}$ S/m). The size, spacing, and drive currents were optimized to account for the steel walls in the lab and to produce a highly accurate field. The mesh of the magnetic-field cage, shown in Fig. \ref{Fig7.1}. For the incident field, we used the Biot-Savart law to compute the fields produced by the six coils in the $x$ direction. This provides for straight segments $C_{j}$ connecting end points $\mathbf{r}_{j}^{(1)\prime }$ and  $\mathbf{r}_{j}^{(2)\prime }$ and forming closed loops $C$ the following expressions for the electric and magnetic fields
\begin{eqnarray}
\mathbf{H}_{1}^{(in)}\left( \mathbf{r}\right)  &=&\sum_{j=1}^{N}\mathbf{H}%
_{1j}^{(in)}\left( \mathbf{r}\right) ,\quad \mathbf{E}_{1}^{(in)}\left( 
\mathbf{r}\right) =\sum_{j=1}^{N}\mathbf{E}_{1j}^{(in)}\left( \mathbf{r}%
\right) ,\quad C=\cup _{j=1}^{N}C_{j},  \label{ex0} \\
\mathbf{H}_{1j}^{(in)}\left( \mathbf{r}\right)  &=&\frac{I_{j}}{4\pi }%
\int_{C_{i}}\frac{d\mathbf{r}^{\prime }\times \left( \mathbf{r-r}^{\prime
	}\right) }{\left| \mathbf{r-r}^{\prime }\right| ^{3}}=\frac{I_{j}}{4\pi }%
\frac{\mathbf{r}_{j}^{(1)}\times \mathbf{r}_{j}^{(2)}}{%
	r_{j}^{(1)}r_{j}^{(2)}+\mathbf{r}_{j}^{(1)}\cdot \mathbf{r}_{j}^{(2)}}\left( 
\frac{1}{r_{j}^{(1)}}+\frac{1}{r_{j}^{(2)}}\right) ,  \notag 
\end{eqnarray}
\begin{eqnarray*}
\mathbf{E}_{1j}^{(in)}\left( \mathbf{r}\right)  &=&\frac{i\omega \mu
	_{1}I_{j}}{4\pi }\int_{C_{j}}\frac{d\mathbf{r}^{\prime }}{\left| \mathbf{r-r}%
	^{\prime }\right| }=\frac{i\omega \mu _{1}I_{j}}{4\pi }\mathbf{e}_{j}\ln
\left| \frac{r_{j}^{(2)}-\mathbf{r}_{j}^{(2)}\cdot \mathbf{e}_{j}}{%
	r_{j}^{(1)}-\mathbf{r}_{j}^{(1)}\cdot \mathbf{e}_{j}}\right| ,  \notag \\
\mathbf{r}_{j}^{(m)} &=&\mathbf{r-r}_{j}^{(m)\prime },\quad
r_{j}^{(m)}=\left| \mathbf{r}_{j}^{(m)}\right| \quad m=1,2,\quad \mathbf{e}%
_{j}=\frac{\mathbf{r}_{j}^{(2)\prime }-\mathbf{r}_{j}^{(1)\prime }}{\left| 
	\mathbf{r}_{j}^{(2)\prime }-\mathbf{r}_{j}^{(1)\prime }\right| }.  \notag
\end{eqnarray*}%
Here $I_{j}$ is the current in the $j$th line element. Figure \ref{Fig7.2} illustrates the incident and scattered magnetic field in the cage at
frequency 60 Hz. As discussed earlier, because the steel walls have a high permeability ($\mu_2/\mu_1 = 100$), this will cause the error to be approximately 100 times higher than if they were non-magnetic. It should be noted that though for such a frequency we have $\delta =2.5$ mm$,$ which is much smaller than the wall thickness ($d=$ 10 cm), and while the skin depth criterion is satisfied, the accuracy of computations for this case is questionable due to large ratio $\mu _{2}/\mu_{1}$, so parameter $\left( \mu _{2}/\mu _{1}\right) \left( \delta /d\right) $ is not small, while $\left( \mu _{2}/\mu _{1}\right) \left( \delta /l\right) $, where $l$ is the characteristic length of the walls (meters) is small.  This is an example of a problem that would benefit from expanding to a second- or third-order approximation, which would allow us to investigate much lower frequencies in the presence of highly magnetic materials.  Such a study is beyond the scope of the present paper and, hopefully, will be conducted in future. Computations with this cage are  shown in Fig. \ref{Fig7.2}. 

\subsubsection{Domain Inversion}
In all of the cases considered above, the conductor occupied a finite domain and
was surrounded by an infinite dielectric.
In practice, however, there are many cases where the opposite situation holds:
the incident field is generated in a finite dielectric surrounded by an infinite conductor. One
such case is an antenna emitting in Earth's waveguide
between the ground and the ionosphere.
To reduce the scale of the problem, we considered a
rectanular box 500 km wide, 500 km long, and 100 km high.
We assumed that both the ground and the ionosphere
were uniform conductors ($\sigma _{2}=1$ mS/m for
the ground and $\sigma _{2}=0.1$ mS/m for the ionosphere) and non-magnetic ($\mu _{2}/\mu _{1}=1$).
At $100$ Hz, the wavelength in air is 3000 km, while the skin depths in the ground and
ionosphere are 1.6 km and 5 km, respectively. Thus, the specified domain size, material properties,
and frequency
satisfy the conditions for which the present approximation is derived.
Note that modeling the ionosphere with a constant, isotropic conductivity is
questionable since the Hall conductivty introduces
anisotropy, especially in the polar regions, where it is strongest \cite{Papadopoulos2011GRL}.
Near the equator, though, the Hall conductivity plays a much lesser role,
so we assumed that the Pedersen and parallel conductivities are of the
same order.
In addition, we assumed that the thickness of the most conductive layer of the ionosphere
is larger than 5 km.
In any event, the case considered here is not designed
for accurate predictions, but rather as an illustration that the present method can be used to consider such problems.

\begin{figure}[htb]
\vspace{-20pt}
\begin{center}
\includegraphics[width=0.9\textwidth, trim=0 0.5in 1.5in 0]{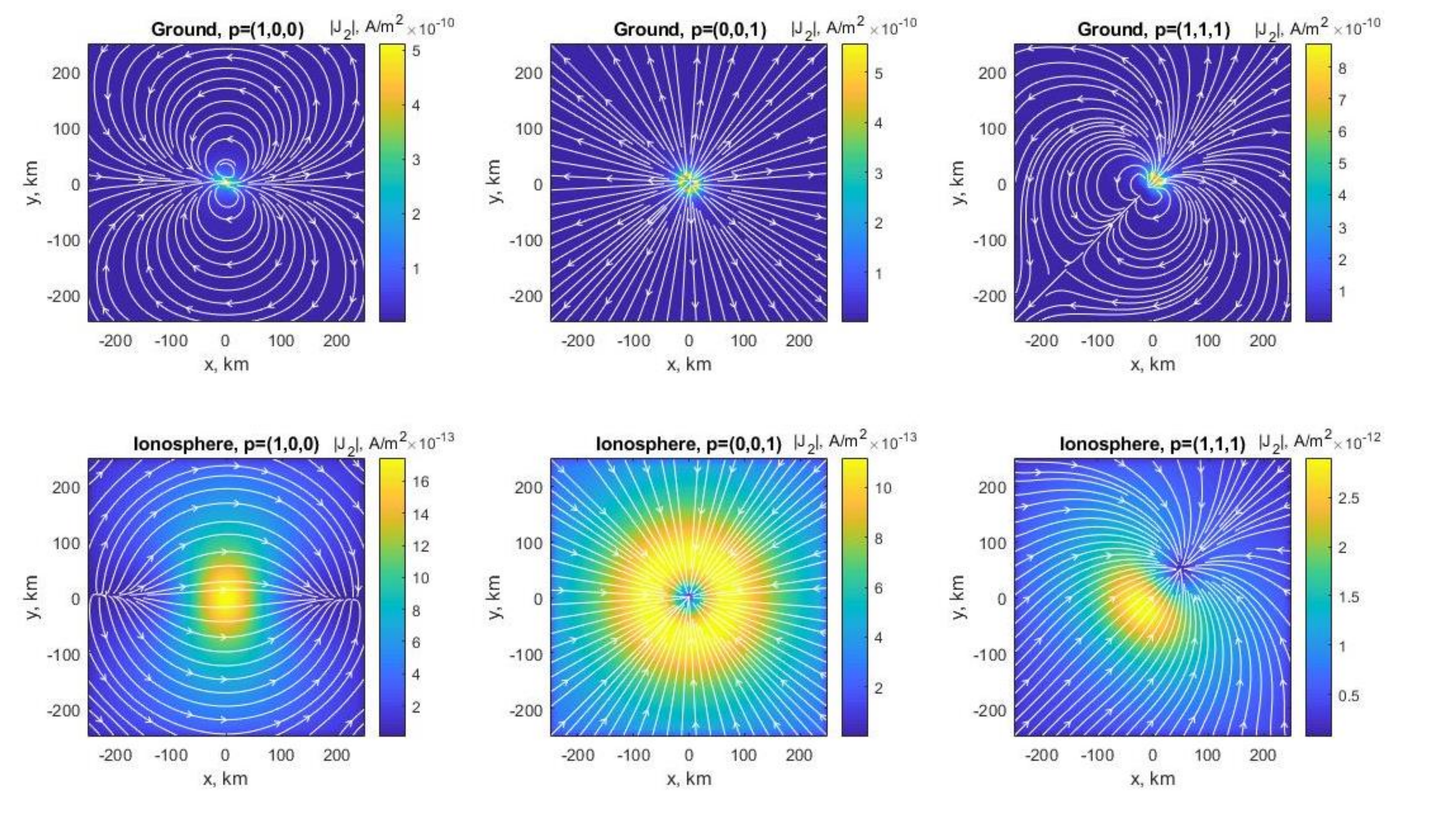}
\end{center}
\caption{The magnitude (colors) and the field lines of the surface currents $%
\left| \mathbf{J}_{2}\right|=\left| \protect\sigma _{2}\mathbf{E}_{2}\right| $ on the ground (the upper
row) and on the lower layer of the ionosphere (the lower row) for different
orientations of a dipole antenna located at \ height 1 km in the Earth's
equatorial region.  The dipole orientation vector ${\bf p}$ is shown for each plot. }
\label{Fig12}
\end{figure}

As the normal to the surface is now directed outside of the domain occupied by
air, the sign near the $1/2$ in Eqs.\ (\ref{e11}) and (\ref{de3}) should be flipped, and that's it!
Specifically, $M+\left(1/2\right)I$ becomes $M-\left(1/2\right)I$, which is now non-singular.
We also note that the boundary conditions on the open sides
of the rectangular domain can be
naturally handled by the accepted single-layer representation of the
potentials. In this case, we simply specify zero charge on those
surfaces.

In the computations, the antenna was modeled as a Hertzian dipole:
\begin{eqnarray}
\mathbf{E}_{1}^{(in)} &=&\mathbf{p}G_{1}\left( \mathbf{r}-\mathbf{r}%
_{s}\right) +\frac{1}{k_{1}^{2}}\nabla \left[ \mathbf{p}\cdot \nabla
G_{1}\left( \mathbf{r}-\mathbf{r}_{s}\right) \right] ,\quad   \label{ex1} \\
\mathbf{H}_{1}^{(in)} &=&\frac{1}{i\omega \mu _{1}}\nabla G_{1}\left( 
\mathbf{r}-\mathbf{r}_{s}\right) \times \mathbf{p,\quad }  \notag \\
G_{1}\left( \mathbf{r}\right) &=&\frac{%
e^{ik_{1}r}}{4\pi r},\quad r=\left| \mathbf{r}\right| , \quad
k_{1} =\frac{\omega }{c_{1}}, \notag
\end{eqnarray}%
where $\mathbf{r}_{s}$ are the coordinates of the dipole, $\mathbf{p}$ is
the dipole moment, and $G_{1}$ is the free-space Green function for the
Helmholtz equation with wavenumber, $k_{1}$, corresponding to the speed of
light, $c_{1}$. The surface of the computational domain was discretized using
7488 triangles. Figure \ref{Fig12} shows the magnitude of the surface
current, $\left| \sigma _{2}\mathbf{E}_{2}\right| $, and the field lines on
the ground and ionosphere surfaces for different values of $\mathbf{p}$. The
antenna is located at the center of the ground surface at a height of $1$ km. The
plotted cases correspond to the dipole, monopole, and mixed type of antennas.

\section{Conclusions}\label{conclusions}

The effects of eddy currents are well known, but their computation can be
challenging due to the need of very high-resolution meshes in full-wave Maxwell
solvers. The method developed in the present work is substantially simpler
than the full-wave solver and produces physically meaningful results.
Comparisons with the exact Mie solution for the sphere show that the computational errors
stay within the error bounds of the method and are mostly determined by the
errors of the approximation rather than the errors of the boundary element
method or the FMM. While the present work provides the framework of how to construct
a general asymptotic expansion, only the zero- and first-order approximations have been explicitly derived. This is sufficient for a number of
practically important problems when solutions for perfect conductors need to
be corrected to account for the imperfectness of real conductors.
However, if the accuracy of the first-order (two-term) approximation of the present work
is insufficient, higher-order approximations could be performed. Such work would be especially useful for highly magnetic materials,
as we observed when the accuracy of the approximation is determined
by the parameter, $\left( \mu _{2}/\mu _{1}\right) \left( \delta /a\right) $, rather than 
$\delta /a$ alone.

\section{Acknowledgment}
This work is supported by Cooperative Research Agreement (W911NF1420118) between the University of Maryland and the Army Research Laboratory, with David Hull and Steven Vinci as Technical monitors.

\end{document}